\documentclass[aps,twocolumn,prb,fleqn]{revtex4}
\usepackage{graphicx,amsfonts,amsbsy}
\usepackage{epsfig,color}
\usepackage{epsf}
\usepackage{graphics}
\usepackage{subfigure}
\def\be{\begin{equation}}
\def\ee{\end{equation}}
\begin{document}

\title{The Magnon Spectrum in the Domain Ferromagnetic State of\\
Antisite Disordered Double Perovskites}
\author{Subrat Kumar Das, Viveka Nand Singh, and Pinaki Majumdar}
\affiliation{Harish-Chandra Research Institute, 
Chhatnag Road, Jhusi, Allahabad, 211 019, India}
\date{5 April 2012}

\begin{abstract}
In their ideal structure, double perovskites like Sr$_2$FeMoO$_6$ 
have alternating Fe and Mo along each cubic axes, and a homogeneous 
ferromagnetic metallic ground state. Imperfect annealing leads to 
the formation of structural domains. The moments on mislocated Fe atoms
that adjoin 
each other across the domain boundary have an antiferromagnetic 
coupling between them. This leads to a peculiar magnetic state, with 
ferromagnetic domains coupled antiferromagnetically. At short distance 
the system exhibits ferromagnetic correlation while at large lengthscales 
the net moment is strongly suppressed due to inter-domain cancellation. 
We provide a detailed description of the spin wave excitations of this 
complex magnetic state, obtained within a $1/S$ expansion, for 
progressively higher degree of mislocation, 
{\it i.e.},  antisite disorder.  At a given 
wavevector the magnons propagate at multiple energies, related, crudely, 
to  `domain confined' modes with which they have large overlap. We provide 
a qualitative understanding of the trend observed with growing antisite 
disorder, and contrast these results to the much broader spectrum that 
one obtains for uncorrelated antisites. 
\end{abstract}
\maketitle

\section{Introduction}

Double perovskite (DP) materials with general formula 
A$_{2}$BB'O$_{6}$ have generated a great deal of interest 
\cite{dp-rev}
both in terms of their basic physics as well as the
possibility of technological applications. 
In particular, Sr$_{2}$FeMoO$_{6}$ (SFMO) shows high 
ferromagnetic $T_c, \sim$ 420K, large electron spin
polarisation (half-metallicity)   
and significant low field magnetoresistance \cite{nat-kob,tom-cryst}. 
 
The ferromagnetic coupling between the $S=5/2$ localized magnetic moments in
SFMO (Fe$^{3+}$ ion, $3d^5$ state) is driven by a 
``double exchange'' mechanism, 
where electrons from Mo delocalise over the Mo-O-Fe network.
The B (Fe) ions order ferromagnetically while the conduction
electrons that mediate the exchange are aligned opposite
to the Fe moments,
leading to a  saturation magnetization of $4 \mu_B$ 
per formula unit in ordered SFMO. However, the large entropy gain from 
disordering promote `antisite disorder' (ASD) 
whereby some B ions occupy the positions of B' ions and {\it vice versa}.

There is clear evidence now that B-B' mislocations 
are not random but spatially correlated 
\cite{asaka-asd-dom,dd-asd-dom}. While ASD suppresses long range 
structural order, electron microscopy \cite{asaka-asd-dom} 
and XAFS \cite{dd-asd-dom} reveal that a high 
degree of short range order survives.
The structural disorder has a direct magnetic impact. 
If two Fe ions adjoin each other the filled shell $d^5$ configuration
leads to antiferromagnetic (AFM) superexchange between them.
The result is
a pattern of structural domains, with each domain internally 
ferromagnetic (FM) while adjoining domains 
are AFM with respect to each other.
This naturally leads to a suppression of the bulk magnetisation
with growing ASD.

Domain structure has been inferred in the low doping manganites as well,
due to competing FM and AFM interactions. Inelastic neutron scattering
in those materials suggest the presence of FM domains in a 
predominantly AFM matrix, and allows a rough estimate of the domain size
\cite{hennion,petit}.  
We aim to provide a similar framework for interpreting
the magnetic state and domain structure in the DP 
from spin wave data.
Our main results are the following.

(i)~We compute the dynamical magnetic structure factor,
 that encodes magnon energy and damping, within a $1/S$
expansion of an effective Heisenberg model chosen to
fit
the electronic model results. (ii)~The magnon data  is
reminiscent of the {\it clean limit} even at maximum
ASD (50\%), where the bulk magnetisation vanishes due to interdomain
cancellation. (iii)~We suggest a rough method for inferring
the domain size from the magnon data and check its 
consistency with the ASD configurations used. (iv)~We
demonstrate that {\it uncorrelated} ASD leads to a much
greater scattering of magnons and a much broader lineshape.
This suggests that in addition to XAFS and microscopy,
neutron scattering would be a sensitive probe of the
nature of disorder in these materials.

The paper is organized as follows: In Sec.II 
we discuss the generation of the structural motif,
the solution of the electronic problem, and
the estimation of  exchanges for an effective Heisenberg model. 
In Sec.III we recapitulate 
the spin-wave formulation for non collinear phases
and present the magnon spectrum obtained for the 
different disordered configurations.
In Sec.IV we discuss the results, attempting to
analyse the magnon spectrum for correlated antisites 
in terms of confined spin wave modes, and contrasting the
result to magnons in an `uncorrelated' antisite background.

\section{Effective magnetic model}

\subsection{Structural motif}

Given the similar location of the
B and B' ions (at the center of the octahedra) the tendency towards
defect formation is more pronounced in the  DP's. This
tendency of mislocation interplays with the inherent B-B' ordering
tendency and creates a spatially 
correlated pattern of antisites \cite{asaka-asd-dom,dd-asd-dom} 
rather than random mislocation. To model this situation
we have used a simple ``lattice-gas''  model \cite{ps-pm-asd}. 
On proper annealing it will go to a long range
ordered   
B, B', B, B'... pattern.
We frustrate this 
by using a short annealing time to 
mimic the situation in the real materials.
We encode the atomic positions by defining a binary 
variable $\eta_{i}$, such that $\eta_{i} =1$ 
when a site has a B ion, and $\eta_{i} =0$ when a site 
has a B' ion. Thus for an ordered 
case we will get $\eta$'s as $1,0,1,0,1,0...$ 
along each cubic axes.
The B-B' patterns that emerge on short annealing are 
characterised by the structural order parameter $S=1-2x$, 
where $x$ is the fraction of B (or B') 
atoms that are  on the wrong sublattice.
We have choosen four disordered families with 
increasing disorder for our study. 
One structural motif each for these families 
is shown in the 
first column of Fig.1, with progressively 
increasing disorder (from top to bottom).
We plot $g({\bf r}_i)=(\eta_i-\frac{1}{2})e^{i\pi(x_i+y_i)}$
as an indicator of structural order.  For a perfectly 
ordered structure $g(\bf{r_i})$ is constant. The pattern along 
the first column are different realisations of ASD   
with $S = 0.98, 0.88, 0.59, 0.17$ (top to bottom). 
We solve the electronic-magnetic problem on these structural motifs.

\subsection{Electronic Hamiltonian}

To study the magnetic order we use the 
electronic-magnetic Hamiltonian that has the usual couplings 
of the ordered double perovskite, and an 
additional antiferromagnetic coupling when two magnetic B 
ions are nearest neighbour (NN). The 
Hamiltonian for the microscopic model is:
\be
H  = H_{loc} \{ \eta\}  + H_{kin} \{ \eta\} + H_{mag} \{ \eta\},
\label{FullHam}
\ee
$H_{loc} \{ \eta\}=\epsilon_{B}\sum_i \eta_i f_{i\sigma}^{\dagger}f_{i\sigma}+
\epsilon_{B'}\sum_i (1 - \eta_i) 
m_{i\sigma}^{\dagger}m_{i\sigma}$ is the onsite term 
where $\epsilon_{B}$ and $\epsilon_{B'}$ 
are level energies, respectively, at the
B and B' sites. Here $f$ is the 
electron operator referring to the magnetic B site and 
$m$ is that of the non-magnetic B' site. 
The NN hopping term is given by 
$H_{kin} \{ \eta\}=-t_{1}\sum_{<ij>\sigma} 
\eta_{i}\eta_{j} f_{i\sigma}^{\dagger}f_{j\sigma} 
-t_{2}\sum_{<ij>\sigma} 
(1-\eta_{i}) (1 - \eta_{j}) m_{i\sigma}^{\dagger}m_{j\sigma}
-t_{3}\sum_{<ij>\sigma} 
(\eta_{i}+\eta_{j}-2\eta_{i}\eta_{j})(f_{i\sigma}^{\dagger}m_{j\sigma}+h.c.)$.
For simplicity we set all the NN hopping 
amplitudes to be same $t_1$=$t_2$=$t_3=t$.  
The magnetic interaction term consists of the 
Hund's coupling $J$ on B sites, and AFM 
superexchange coupling $J^{AF}$ between two NN magnetic B sites. Thus,
$H_{mag} \{ \eta\}=
J\sum_i \eta_i {\bf S}_{i}.f_{i\alpha}^{\dagger}\vec{\sigma}_{\alpha\beta}f_{i\beta} 
+ J^{AF} \sum_{\langle ij \rangle} \eta_i \eta_j {\bf S}_i.{\bf S}_j $. 
Here ${\bf S}_i$ is the classical core spin 
on the B site at ${\bf r}_i$ with $\vert {\bf S}_i \vert = 1$.
We take $J/t \gg 1$ with $J>0$ and  
$J^{AF}\vert {\bf S} \vert^2/t = 0.08$, based on the 
T$_N$ scale in SrFeO$_3$. We have ignored 
orbital degeneracy, coulomb effects, {\it etc}, to
focus on the essential magnetic model on the
disordered structure.
We will use a two dimensional model because it already
captures the qualitative physics while allowing ease of visualisation
and access large system size. The formulation readily carries
over to three dimensions as well.

We have used a real space exact diagonalisation 
based Monte Carlo method involving a 
traveling cluster approximation (TCA)\cite{tca} to 
anneal the spin-fermion system towards its ground state
in the disordered background. 

\begin{figure}[t]
\centerline{
\includegraphics[width=2.7cm,height=2.7cm,angle=0,clip=true]{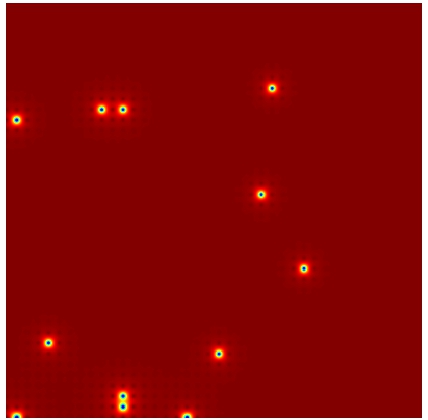}
\includegraphics[width=2.7cm,height=2.7cm,angle=0,clip=true]{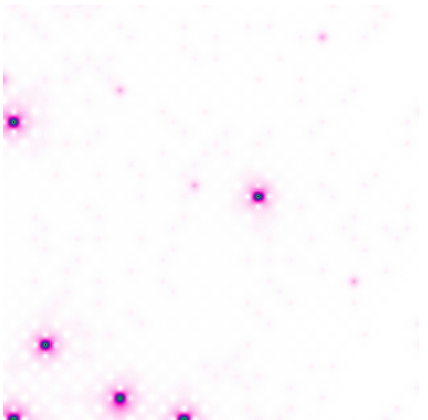}
\includegraphics[width=2.7cm,height=2.7cm,angle=0,clip=true]{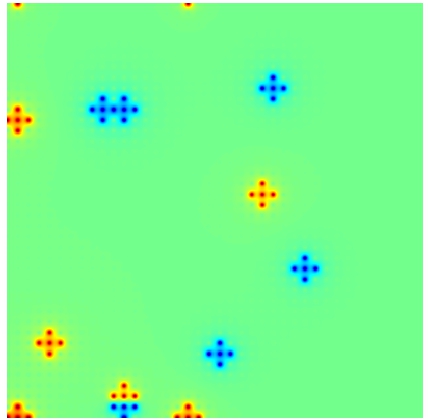}
}
\vspace{.2cm}
\centerline{
\includegraphics[width=2.7cm,height=2.7cm,angle=0,clip=true]{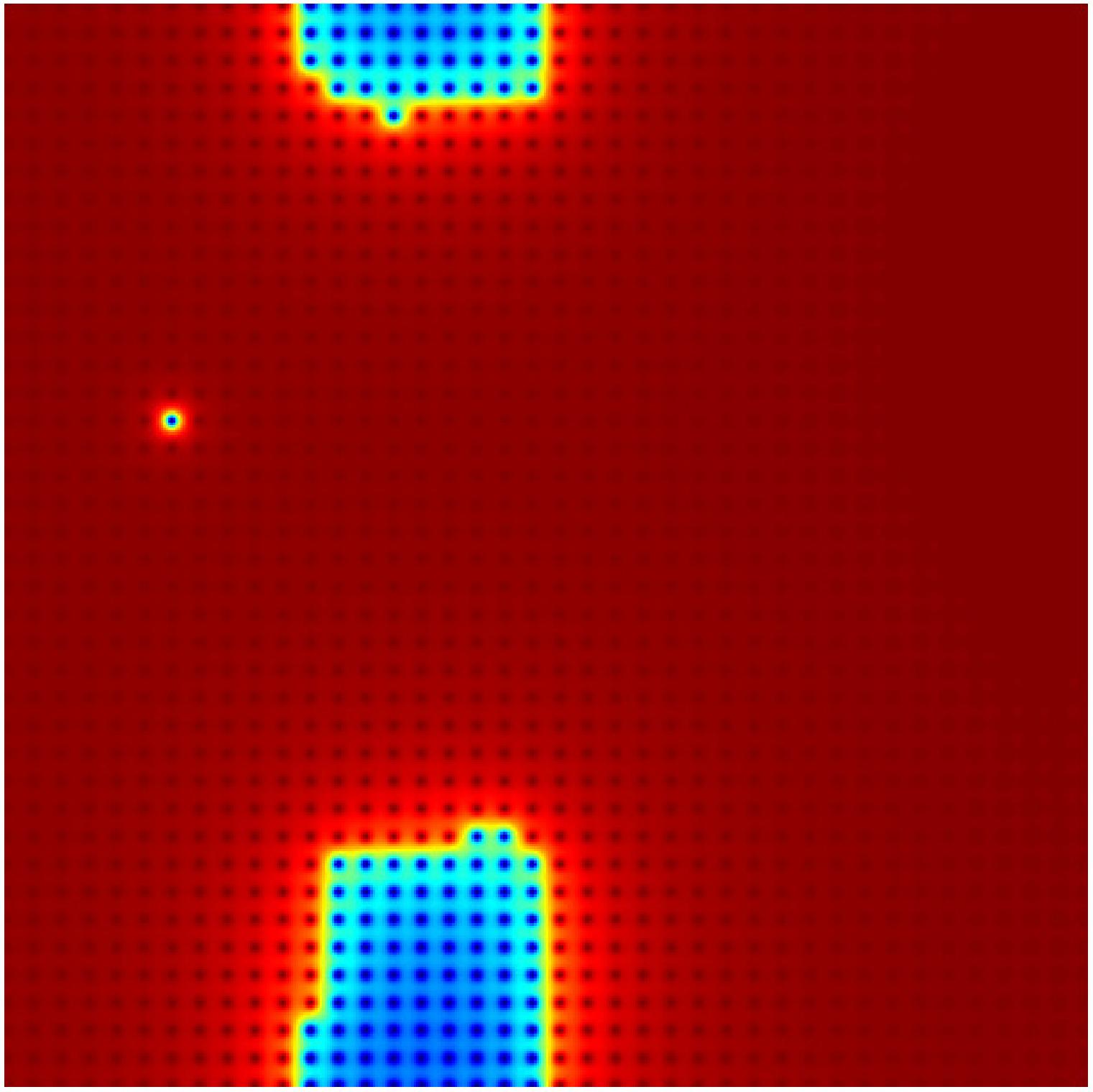}
\includegraphics[width=2.7cm,height=2.7cm,angle=0,clip=true]{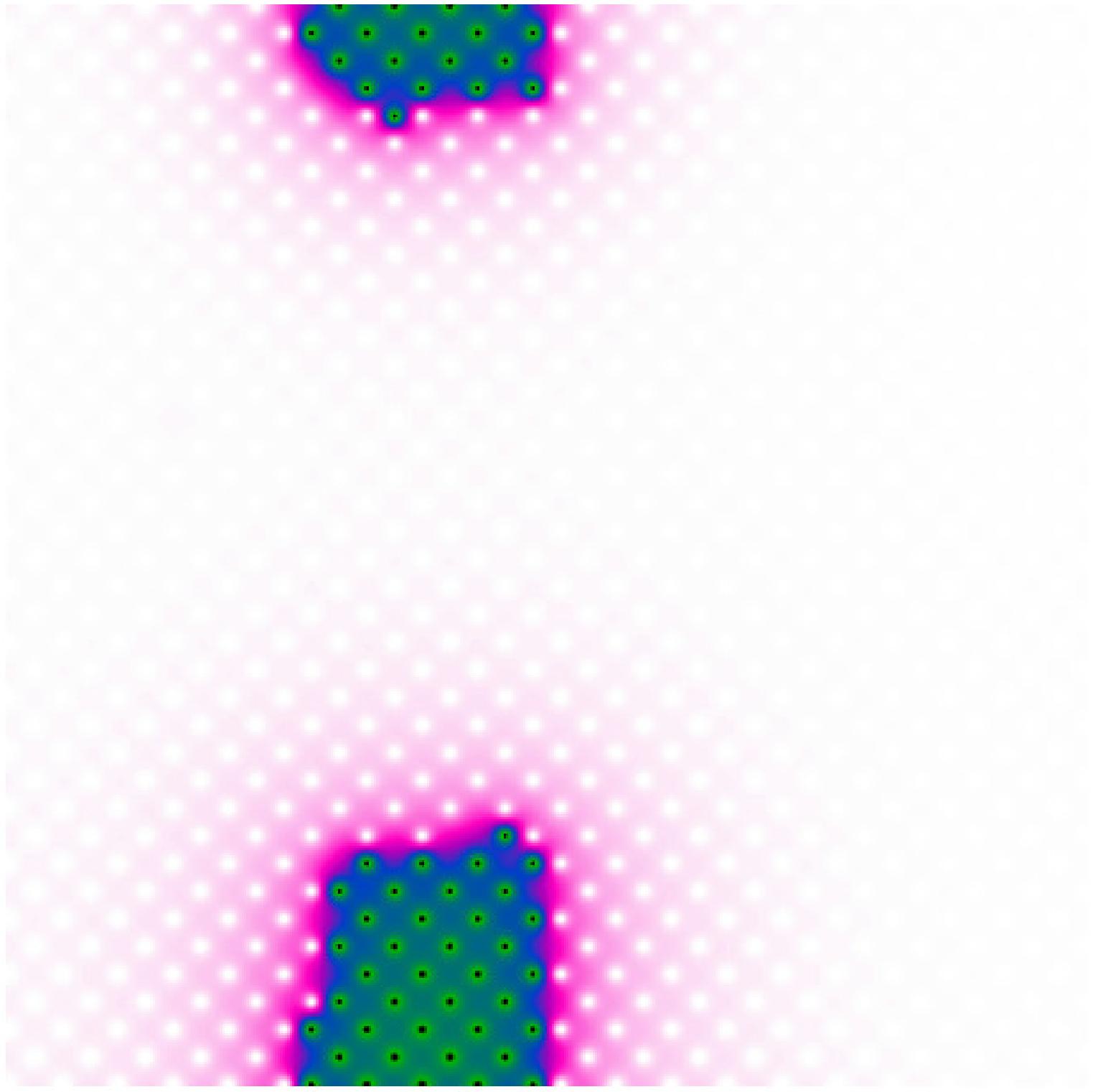}
\includegraphics[width=2.7cm,height=2.7cm,angle=0,clip=true]{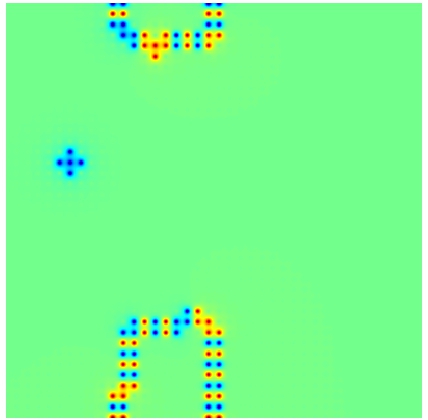}
}
\vspace{.2cm}
\centerline{
\includegraphics[width=2.7cm,height=2.7cm,angle=0,clip=true]{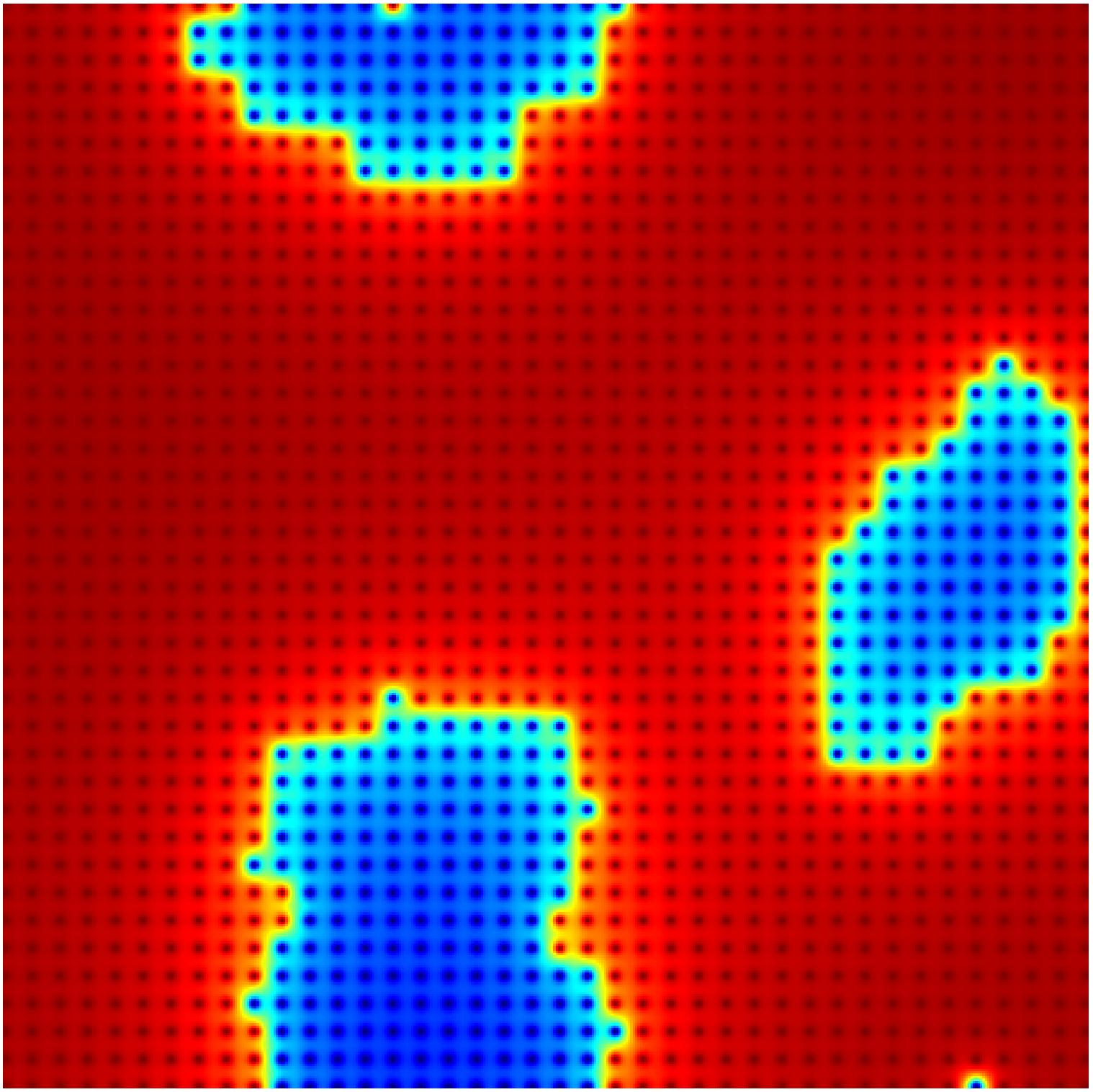}
\includegraphics[width=2.7cm,height=2.7cm,angle=0,clip=true]{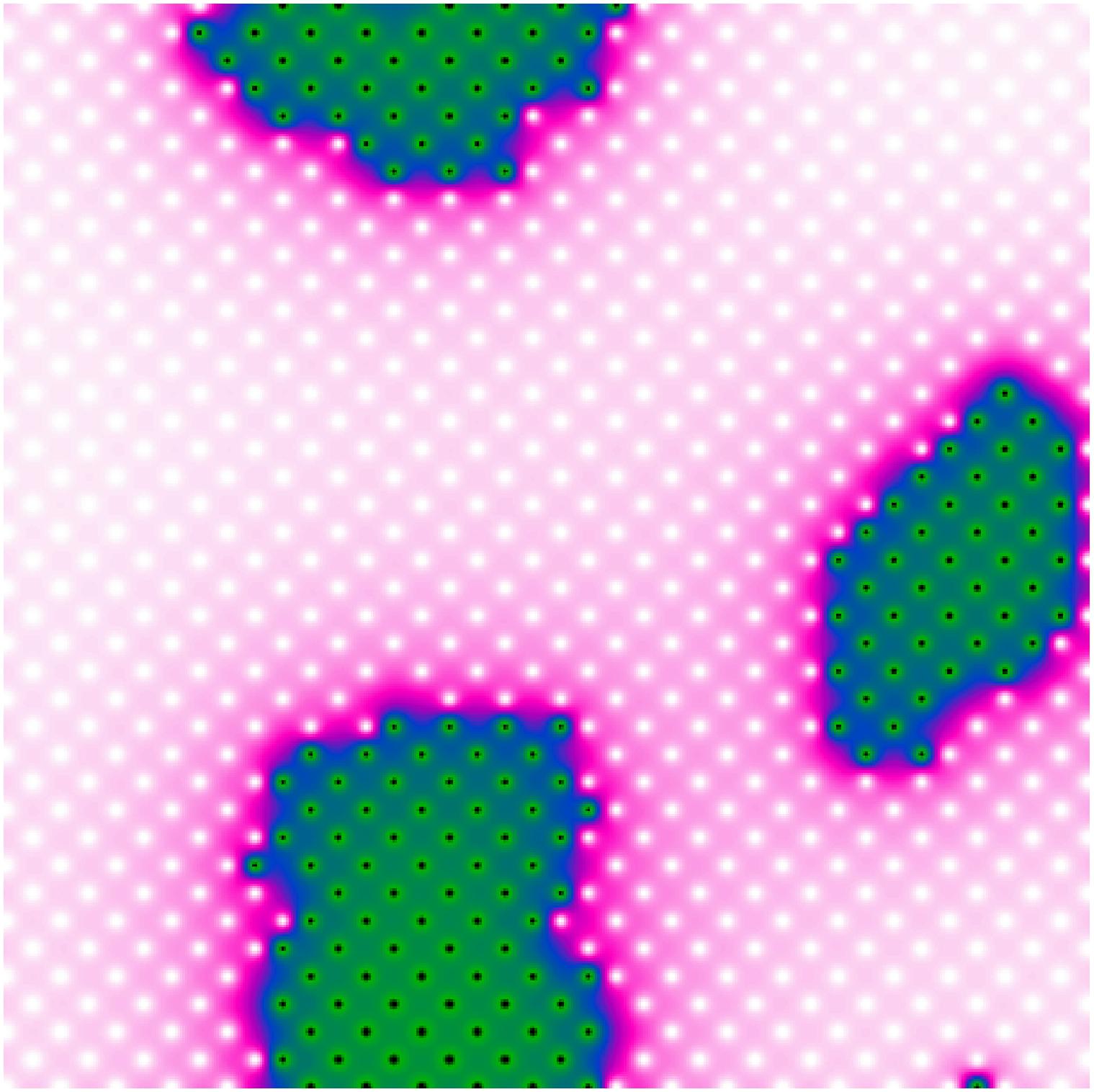}
\includegraphics[width=2.7cm,height=2.7cm,angle=0,clip=true]{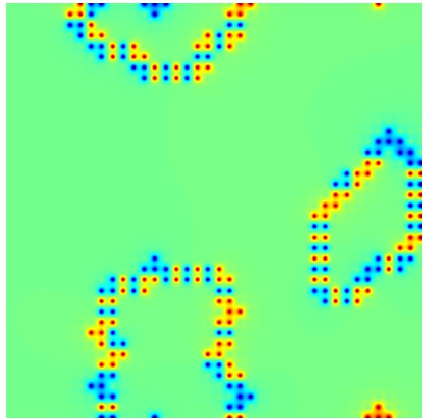}
}
\vspace{.2cm}
\centerline{
\includegraphics[width=2.7cm,height=2.7cm,angle=0,clip=true]{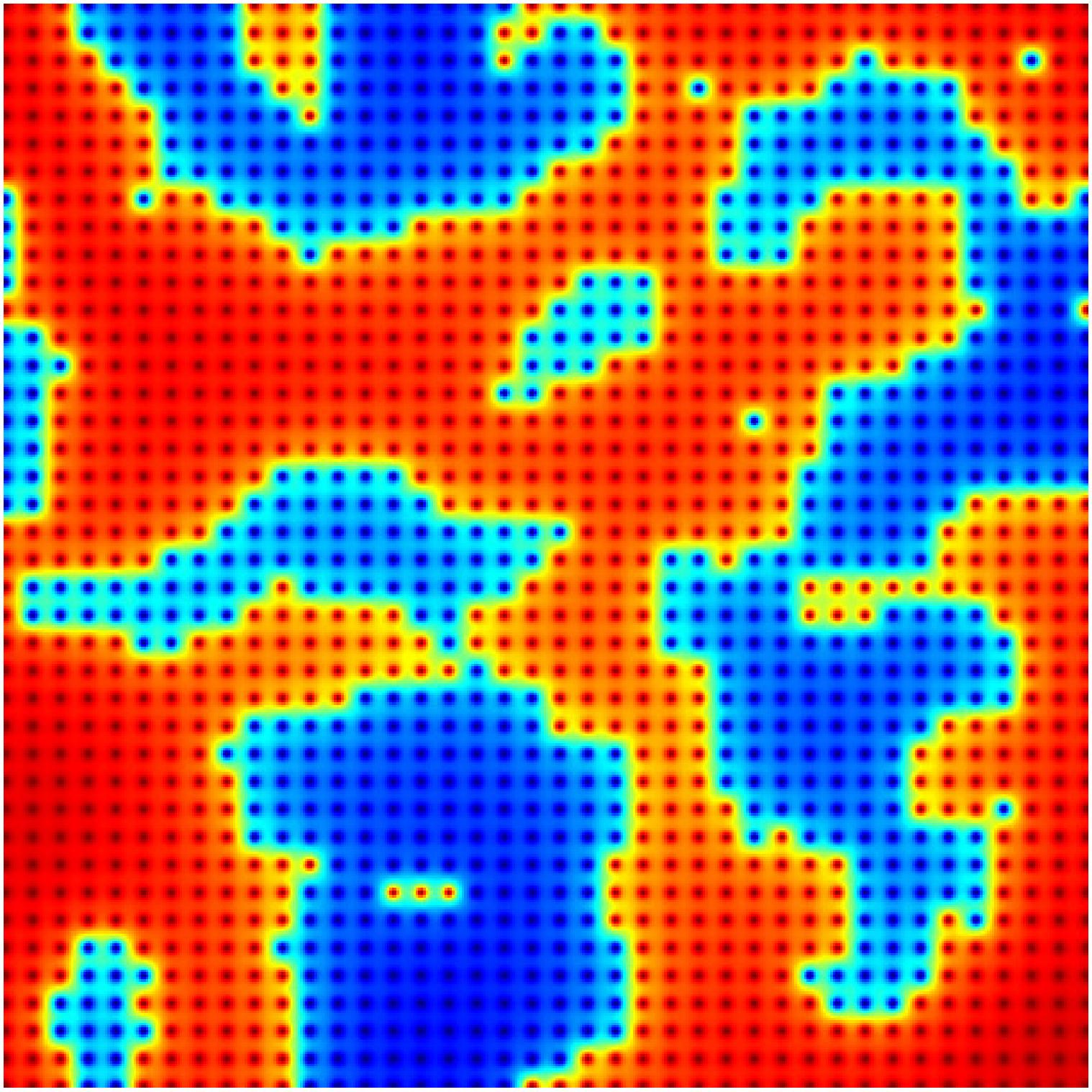}
\includegraphics[width=2.7cm,height=2.7cm,angle=0,clip=true]{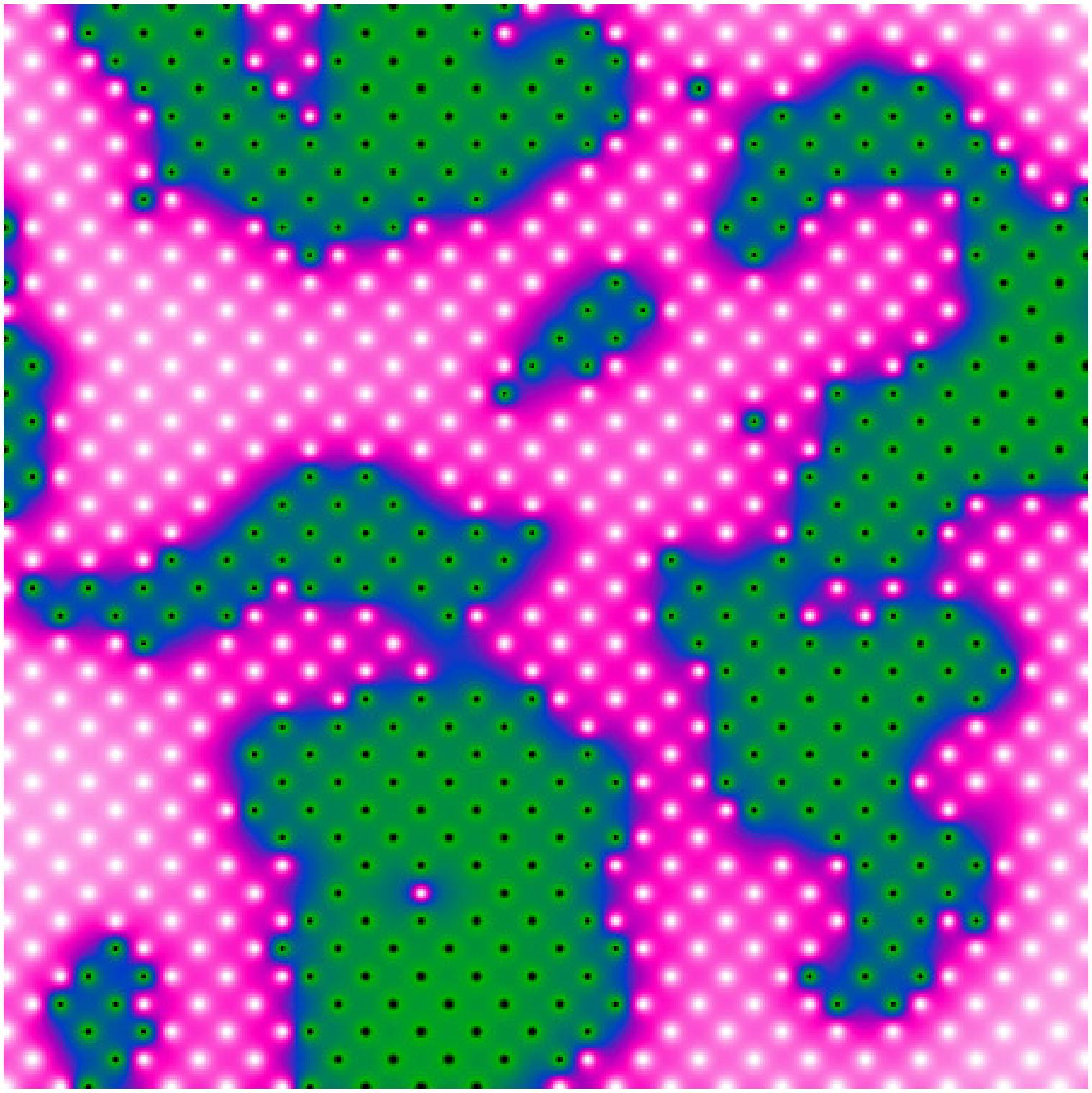}
\includegraphics[width=2.7cm,height=2.7cm,angle=0,clip=true]{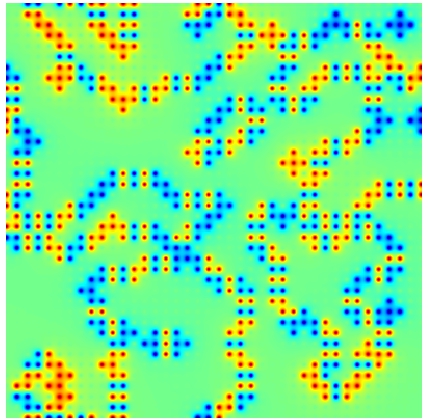}
}
\caption{(Colour online) 
First column contains the structural motif for four disordered 
families with progressively 
increasing disorder (from top to bottom). 
We plot the $g(\bf{r_i})=(\eta_i-\frac{1}{2})e^{i\pi(x_i+y_i)}$ . 
We denote the configurations as C1, C2, C3, C4 
and the corresponding   
structural order parameter has values $S = 0.98,~0.88,~0.59,~0.17$ 
from top to bottom. 
Second column shows the ground state spin overlap factor, 
$h_i=\bf{S_0}.\bf{S_i}$, where $\bf{S_0}$ 
is the left-lower-corner spin in the lattice.
In the third column, we have shown the corresponding 
NN bond configurations. Here we have three 
different type of NN bonding present between 
B-B, B'-B' and B-B', represented by 
colours red, blue and green respectively in the plot.
Lattice size is $40 \times 40$.
}
\label{fig:str}
\end{figure}

Annealing the electron-spin system 
down to low temperature on a given structural 
motif leads to the magnetic ground states shown in
the middle column of Fig.1.  
We plot the  spin overlap factor, 
$h_i=\bf{S_0}.\bf{S_i}$, where $\bf{S_0}$ 
is the left-lower-corner spin in the lattice.
The comparison of the first and second columns in Fig.1 
indicate that the  structural 
and magnetic domains coincide with each other. 
The third column of Fig.1 shows the NN 
structural partners. We have three 
possibilities: 
B-B, B'-B' and B-B', represented by 
colours red, blue and green respectively in the plot. 

\subsection{Effective Heisenberg Hamiltonian}

Considering the difficulty in doing a spin-wave analysis 
on the full electronic-magnetic Hamiltonian (Eq. \ref{FullHam}), 
we assume that the spin dynamics can be described 
by an $\it effective$ Heisenberg model
\be
H_{eff} = \sum_{\{ ij \}}J_{ij}\;{\bf S}_i.{\bf S}_j
\label{HeisenbergHam}
\ee
where $\{\}$ represents the set of 
NN and next nearest neighbour (NNN) sites. 
$J_{ij}$ is the effective coupling (FM/AFM) 
between the local moments at ${\bf r}_i$ and ${\bf r}_j$ sites.
In our two dimensional ASD configurations 
$J_{F}$ operates between two local moments 
when they are at the NNN position
and $J_{AF}$ is active when the moments are at the 
NN position (a B-O-B arrangement). We have estimated the effective 
coupling $J_{F}$ and $J_{AF}$ as follows. For getting the 
FM coupling ($J_{F}$) we have considered the ordered double 
perovskite structure.
We calculated the order parameter,
{\it i.e}, the magnetic structure 
factor S($\bf{k}$) at ${\bf {k}}=(0,0)$, as a 
function of temperature for the full electronic Hamiltonian 
(Eq. \ref{FullHam}) using Monte Carlo simulation. 
We then repeated the same procedure for the NNN FM 
Heisenberg Hamiltonian, defined 
on only the magnetic sites of the double perovskite.
We found that for $J_{F}/t=-0.04$, two results matches very well. 

\begin{figure}[t]
\centerline{
\includegraphics[width=6.4cm,height=4.5cm,angle=0,clip=true]{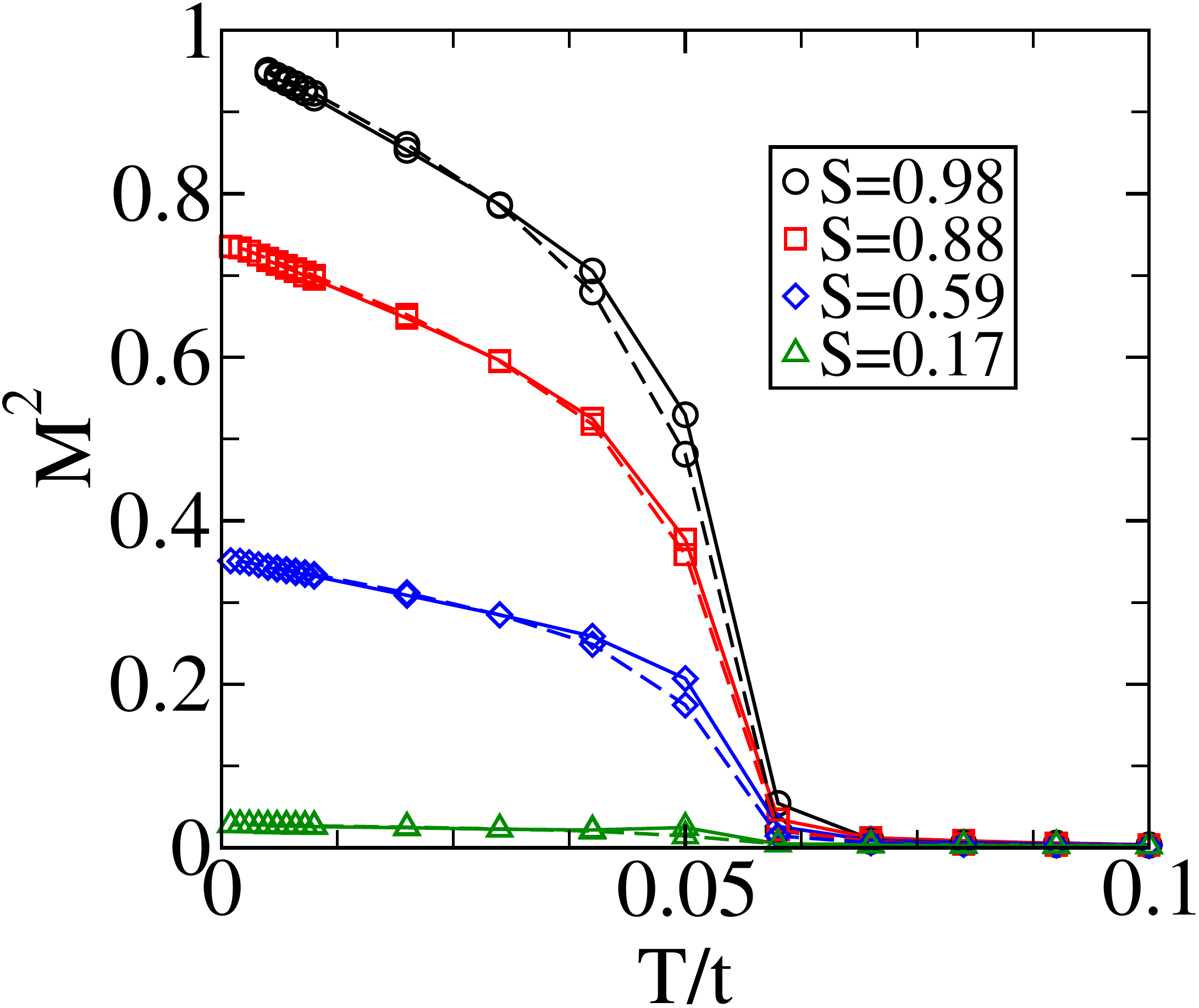}
}
\caption{(Colour online) Comparison between the evolution of the 
spin structure factor S($\bf k$) at ${\bf k} = (0,0)$ 
with temperature for the spin configurations of various disorder 
families (from top to bottom) C1, C2, C3 and C4 obtained 
from the full electronic Hamiltonian with 
$J^{AF}S^2/t = 0.08$ and the effective Heisenberg model 
with $J_{F}/t=-0.04$ and $J_{AF}/t=0.065$. 
Lattice size is $40 \times 40$. 
}
\end{figure}

In order to get the AFM coupling we considered the 
ordered perovskite where both the B and B' site carry a 
magnetic moment (mimicking SrFeO$_3$) and 
computed its AFM structure                                                   
factor peak {\bf k}=($\pi,\pi$). This model
involves both electronic kinetic energy and Fe-Fe
superexchange.  
We find that the result can be modelled via
a Heisenberg model with $J_{AF}/t=0.065$. 

Using the couplings inferred from these limiting cases,
 $J_{F}/t=-0.04$ and $J_{AF}/t=0.065$, 
we studied the bond disordered  Heisenberg   
model for the  antisite disordered DP magnet.
We compared the FM structure factor peak 
S($\bf {k}$) at ${\bf k}$=(0,0) obtained from the disordered 
Heisenberg model with that from 
the full electronic Hamiltonian (Eq. \ref{FullHam}). 
The Heisenberg result for the FM structure 
factor S(0,0) as a function of temperature matches very well,
Fig.2,  with the electronic Hamiltonian result for all ASD 
configurations. This gives us confidence in the usefulness of the
Heisenberg model for spin dynamics.

\section{Spin dynamics}

\subsection{Spin-Wave Excitation} 

In this section we use the spin rotation technique 
\cite{fishmanJPCM21} to evaluate the spin-wave modes and 
dynamic structure factor at zero temperature.
The effective Heisenberg model (Eq. \ref{HeisenbergHam}) 
can be cast in a form useful for spin wave analysis 
by defining a {\it local frame} at each site so that the spins
point along the $+z$ direction in the ground state.
We can use
${\bar{\bf S}}_i = U_i{\bf S}_i$, 
where ${\bar{\bf S}}_i$ points along its 
local $z-$axis in the 
classical limit. The unitary rotation matrix 
$U_i$ for site ${\bf r}_i$ is given by 
\be 
U_i = 
 \left| \begin{array}{ccc}
\cos(\theta_i)\cos(\psi_i) & \cos(\theta_i)\sin(\psi_i)  
& -\sin(\theta_i) \\
-\sin(\psi_i) & \cos(\psi_i)& 0 \\
\sin(\theta_i)\cos(\psi_i) & \sin(\theta_i)\sin(\psi_i) 
& \cos(\theta_i) \end{array} \right|,
\label{rotationmatrix}
\ee
where $\theta$ and $\psi$ are the Euler rotation 
angles. Now one can write the 
generalized Hamiltonian
\be
H_{eff} = \sum_{\{ ij \}}
J_{ij}{\bar{\bf S}}_i.F_{ij}{\bar{\bf S}}_j,  
\label{rotatedHam}
\ee
where $F_{ij}= U_iU_j$ is the overall rotation 
from one reference frame to another 
and its elements $F_{ij}^{\alpha\beta}$ 
can be obtained from Eq. (\ref{rotationmatrix}).   

Applying the approximate Holstein-Primakoff (HP) 
transformation in the large $S$ limit 
the spin operators in the local reference frame 
become: $\bar{S}_i^+ = \sqrt{2S}\:b_{i}$, 
$\bar{S}_i^- = \sqrt{2S}\:b_{i}^{\dag}$ and 
$\bar{S}_i^z = S -b_{i}^{\dag} b_{i}$, where 
$b_i$ and $b_i^{\dag}$ are the boson (magnon) 
annihilation and creation operators respectively. 
Only retaining the quadratic terms in $b$ and 
$b^{\dag}$, which describe the dynamics of the 
non-interacting magnons and neglect magnon 
interaction terms of order $1/S$, the 
generalized Hamiltonian (Eq. \ref{rotatedHam}) reduces to 
\be
{\cal H} = \sum_{\{ ij \}}
[{\cal J}_{ij}(G_{ij}^1 b_i^{\dag}b_j + 
G_{ij}^2 b_ib_j + h.c.) + 
f_{ij}(b_i^{\dag}b_i + b_j^{\dag}b_j)],
\label{quaraticdHam}
\ee
where ${\cal J}_{ij} =SJ_{ij}/2$, $f_{ij} = 
-SJ_{ij}F_{ij}^{zz}$ and the rotation coefficients 
$G^{{}^1_2}= (F_{ij}^{xx} \pm F_{ij}^{yy}) - i(F_{ij}^{xy} \mp F_{ij}^{yx})$. 
The Hamiltonian (\ref{quaraticdHam}) is diagonalized by the transfermation  
\be
b_i = \sum_n (u_n^i c_n + v_n^{i^{*}} c_n^{\dag}),
\label{BogoliubovTrans}
\ee
where $c^{\dag}$ and $c$ are the quasiparticle operators. 
$u$ and $v$, which satisfy $\sum_n (u_n^i u_n^{j^{*}} - 
v_n^{i^{*}} v_n^j) = \delta_{ij}$ 
ensuring the bosonic character of the quasiparticles are obtained from  
\be
\left( \begin{array}{cc}
A_{ij} & B_{ij}^* \\
B_{ij} & A_{ij}^*  
\end{array} \right)
\left( \begin{array}{c}
u_n^j\\
v_n^j  
\end{array} \right) = \omega_n
\left( \begin{array}{cc}
\delta_{ij} & 0 \\
0 & -\delta_{ij}  
\end{array} \right)
\left( \begin{array}{c}
u_n^j\\
v_n^j  
\end{array} \right), 
\label{BogoliubovEqs}
\ee
where $A_{ij} = {\cal J}_{ij}(G_{ij}^1 + 
G_{ji}^{1^{*}}) + \epsilon_i\delta_{ij}$, 
$B_{ij} = {\cal J}_{ij}(G_{ij}^2 + G_{ji}^2)$ and
$\epsilon_i = \sum_j(f_{ij} + f_{ji})$.  
Now the spin-spin correlation function can be evaluated using the 
magnon energies and wavefunctions  
obtained from Eq. (\ref{BogoliubovEqs}), where the 
excitation eigenvalues $\omega_n \geq 0$.

\subsection{Dynamical Structure Factor}

A neutron scattering experiment measures the spin-spin correlation 
function in Fourier and frequency space 
$S({\bf k},\omega)$ to describe the spin dynamics of the magnetic 
systems on an atomic scale. 
From ${\bf S}_i =  U_i^{-1}{\bar{\bf S}}_i$ 
one can express
$
S_i^{\alpha} = \sum_{\mu} U_i^{{\mu \alpha}^*} {\bar {S}_i^{\mu}},
\label{Sialpha}
$ 
where $\alpha$ and $\mu$ represents the $x$, $y$, and $z$ components. 
Now applying the approximate HP transformation to the 
rotated spins one can write 
\be
S_i^{\beta} = p_i^{\beta}b_i + 
q_i^{\beta}b_i^{\dag} + r_i^{\beta}(S - b_i^{\dag}b_i),
\label{Sibeta}
\ee 
where $\beta = +$, $-$ and $z$, and $p,q$ and $r$ are the 
rotation coefficients (given in the Appendix).

Putting Eq. (\ref{BogoliubovTrans}) in (\ref{Sibeta}) 
the space time spin-spin correlation function 
can be written as  
\be
S_i^{\alpha}(t) S_j^{\beta}(0) = 
\sum_{mn}[A_{{}^{mn}_{ij}}^{\alpha \beta} 
c_m^{\dag}(t) c_n(0) + 
B_{{}^{mn}_{ij}}^{\alpha \beta} c_m(t) c_n^{\dag}(0)], 
\label{SiAlphaSjBeta}
\ee
where the coefficients A and B are expressed in the Appendix.
In Fourier and frequency space 
\be
S^{\alpha , \beta}({\bf k},\omega) = 
\frac{1}{N}\int dt e^{- i\omega t} \sum_{ij} 
e^{i{\bf k}.({\bf r}_i - {\bf r}_j)} 
\langle S_i^{\alpha}(t)S_j^{\beta}(0)  \rangle. 
\label{CorrFunction}
\ee
and the total spin-spin correlation function
\begin{eqnarray}
S({\bf k},\omega) &=& \frac{1}{2}
[S^{+,-}({\bf k},\omega)+S^{-,+}({\bf k},\omega)] 
+ S^{z,z}({\bf k},\omega) \nonumber\\
&=& \sum_{l} W_{\bf k}^{l} 
\delta (\omega - \omega_{l}), \nonumber 
\end{eqnarray}
where the coefficient of the delta function 
\be
W_{\bf k}^{l}=\frac{1}{N}\sum_{ij} 
{\cal B}_{ij}^l e^{i{\bf k}.({\bf r}_i - {\bf r}_j)}  
\label{weight}
\ee
is the SW weight with ${\cal B}_{ij}^l = 
\frac{1}{2}(B_{{}^{ll}_{ij}}^{+-} + 
B_{{}^{ll}_{ij}}^{-+}) + B_{{}^{ll}_{ij}}^{zz}$. 
$W_{\bf k}^{l}$ is observed as the intensity of 
magnon spectrum in the neutron scattering experiment.

\begin{figure}[b]
\centerline{
\includegraphics[width=4.0cm,height=4cm,angle=0,clip=true]{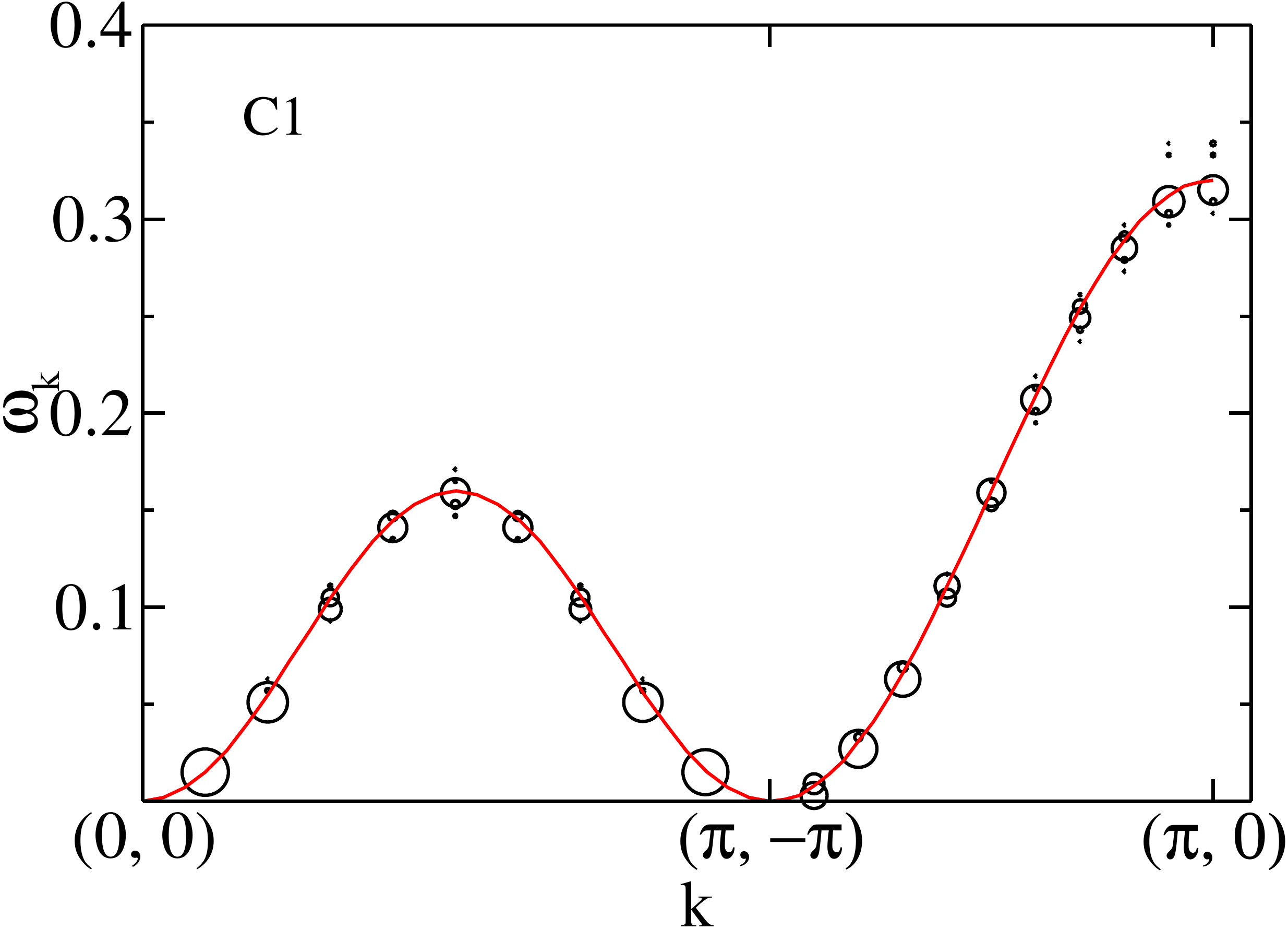}
\includegraphics[width=4.0cm,height=4cm,angle=0,clip=true]{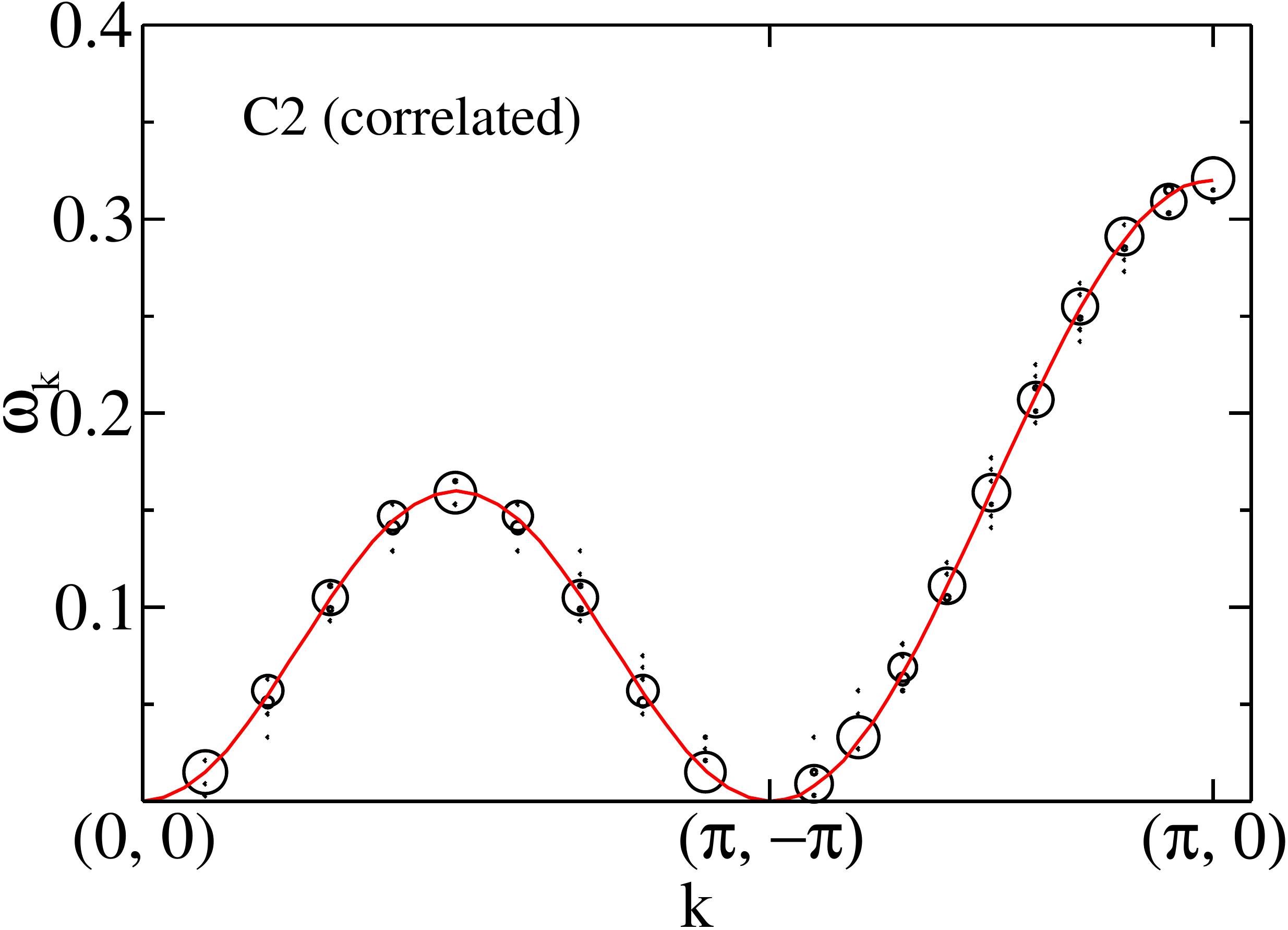}
}
\centerline{
\includegraphics[width=4.0cm,height=4cm,angle=0,clip=true]{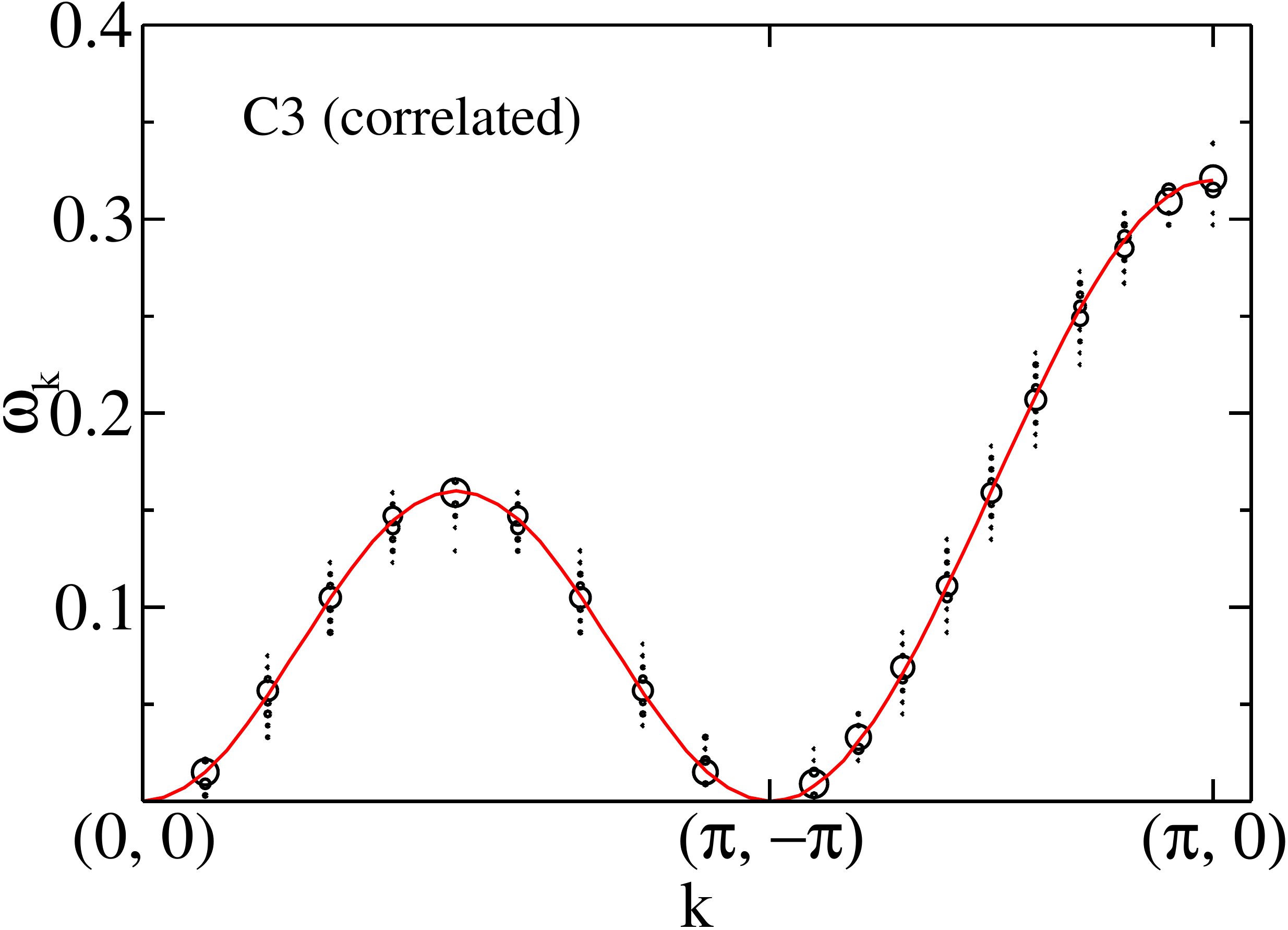}
\includegraphics[width=4.0cm,height=4cm,angle=0,clip=true]{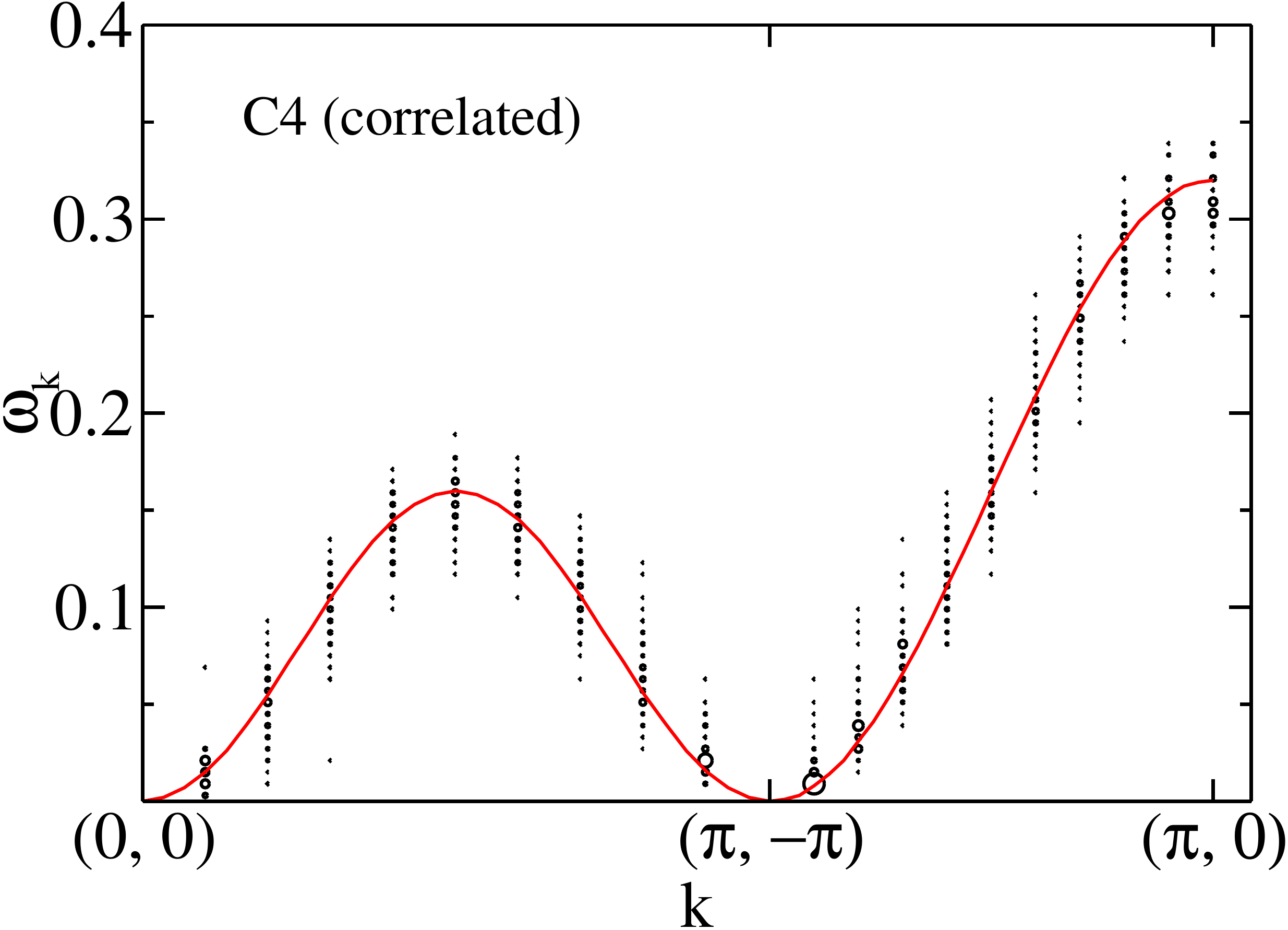}
}
\caption{ (Colour online)  Spin-wave spectrum along main symmetry
directions of the Brillouin zone for spin configurations C1, C2, C3 and C4
 ($x$ = 0.01, 0.11, 0.21 and 0.41 respectively).  
shown in Fig.1 
With increasing
ASD from C1 to C4 the spectrum becomes broader for a fixed value of
momentum ${\bf k}$. Here $J_{F}=-0.04$, $J_{AF}=0.065$, and lattice size is
$40 \times 40$.}
\label{fig:swe}
\end{figure}
\begin{figure}[t]
\centerline{
\includegraphics[width=5.5cm,height=4.5cm,angle=0,clip=true]{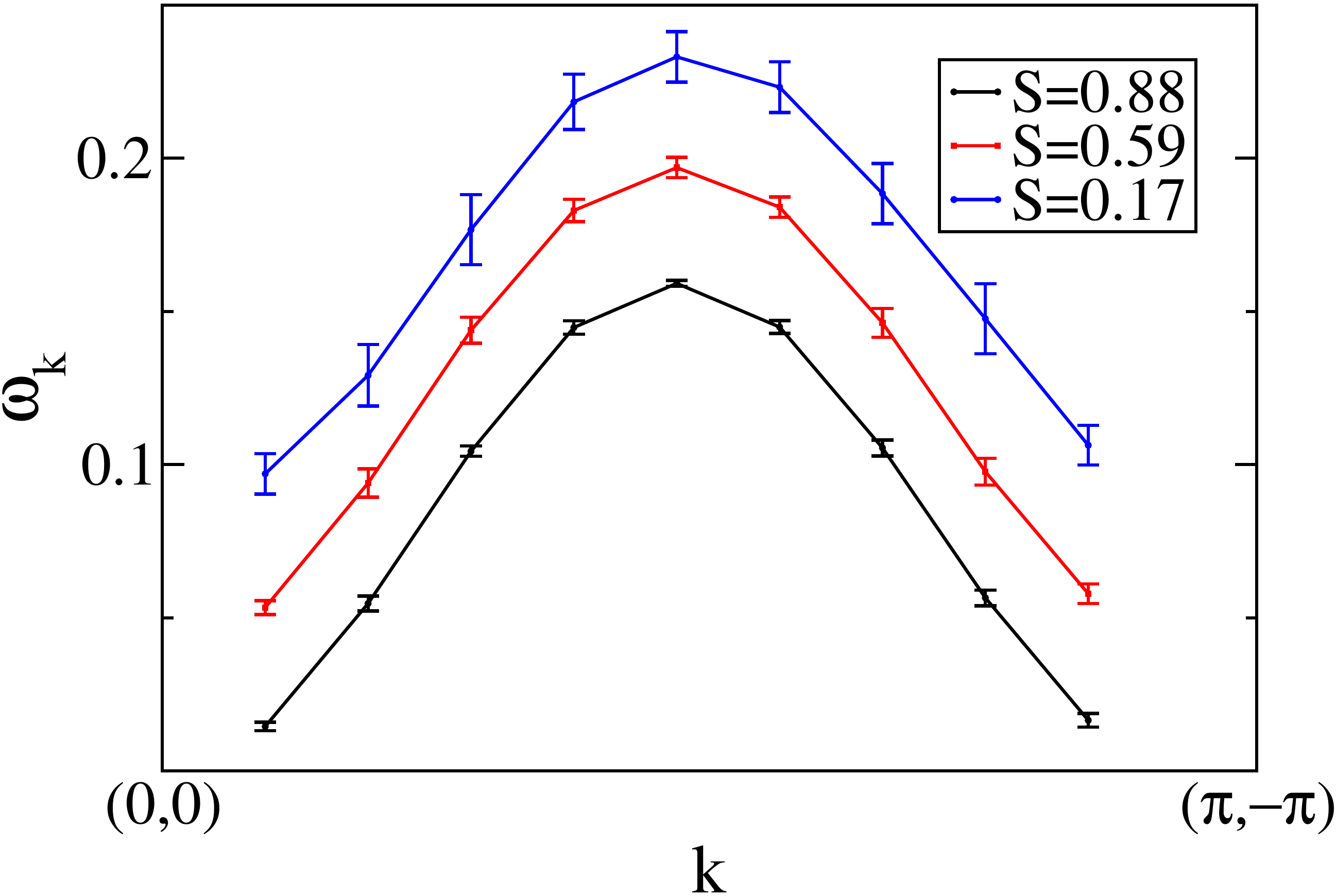}
}
\caption{ (Colour online)  
Mean spin wave energy ${\bar \omega}_{\bf k}$ (dots) and 
spin-wave width $\Delta \omega_{\bf k}$ (bars), defined in the 
text, 
for the correlated antisite configurations C2-C4. The curves are
vertically shifted for clarity.
}
\end{figure}

\section{Results and Discussion}

We start by presenting the results for magnons
in the configurations C1-C4 shown in Fig.1 and 
then move to an analysis of the linewidth,
the estimation of domain size, and the contrast with
uncorrelated disorder.

\subsection{Results for AF coupled domains}

Fig.3 shows the magnon spectra of 
C1-C4 with obtained from the Heisenberg model with
the FM and AFM couplings discussed earlier. 
In a model with only FM couplings, {\it i.e.}, no
disorder, we would have obtained only the red curve,
$\omega^0_{\bf k}$, 
for propagating magnons.
The striking feature in all these panels is
how closely the mean energy of the magnons follow
$\omega^0_{\bf k}$
despite the large degree of mislocation in C2 and C3
and maximal disorder ($ x \sim 0.5$) in C4 (refer to the spatial
plots in Fig.1).
The broadening, although noticeable in C4, does
not obscure the basic dispersion. 

Fig.4 quantifies the mean energy and broadening by
computing:
$$
{\bar \omega}_{\bf k} = \int~S({\bf k}, \omega) \omega d \omega
$$
$$ 
[ {\Delta \omega_{\bf k}} ]^2 =
[\int~S({\bf k}, \omega) \omega^2 d \omega] - {\bar \omega}^2_{\bf k}
$$
We have shown these two quantities for the C2-C4  structures
in Fig.1. The ${\bar \omega}_{\bf k}$
have been vertically shifted for clarity and the ${\Delta \omega_{\bf k}}$
are superposed as `error bars' on these. It is clear that even in
the most disordered sample (C4), where the mislocation $x \sim 0.4$,
the broadening is only a small fraction of the magnon energy. 
This will be an indicator when we discuss spin waves in an
uncorrelated disorder background.

\begin{figure}[b]
\centerline{
\includegraphics[width=4.0cm,height=4cm,angle=0,clip=true]{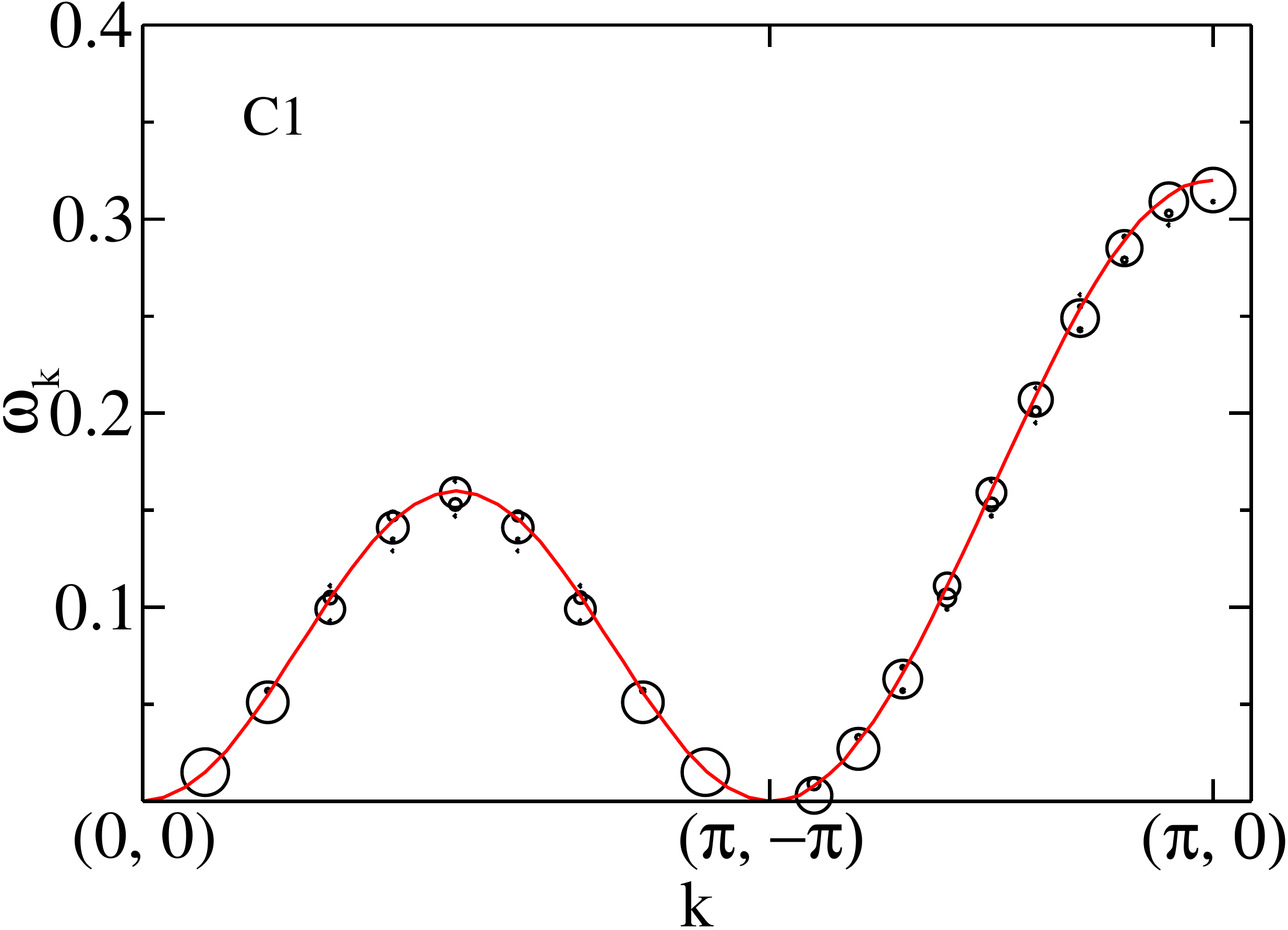}
\includegraphics[width=4.0cm,height=4cm,angle=0,clip=true]{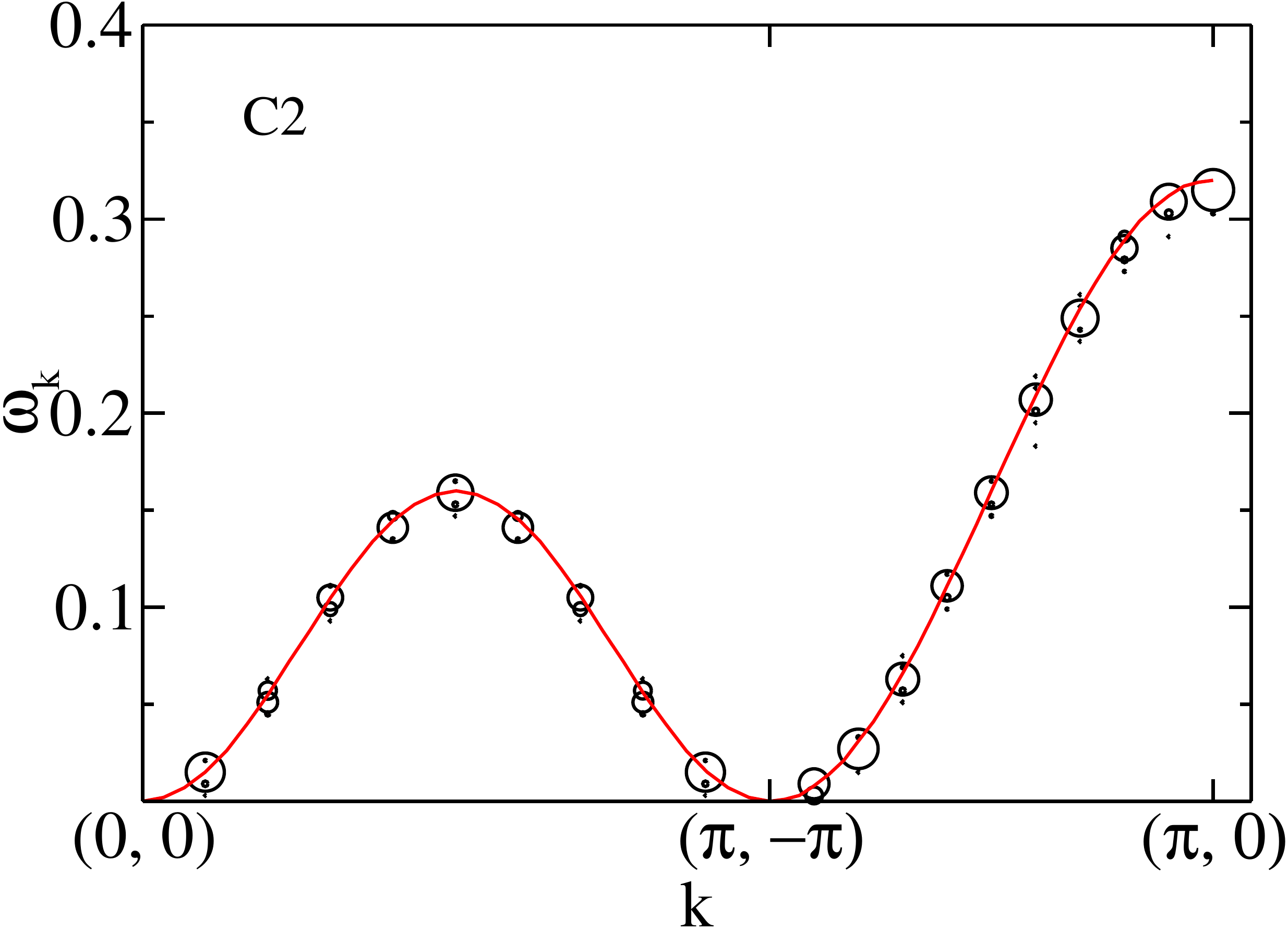}
}
\centerline{
\includegraphics[width=4.0cm,height=4cm,angle=0,clip=true]{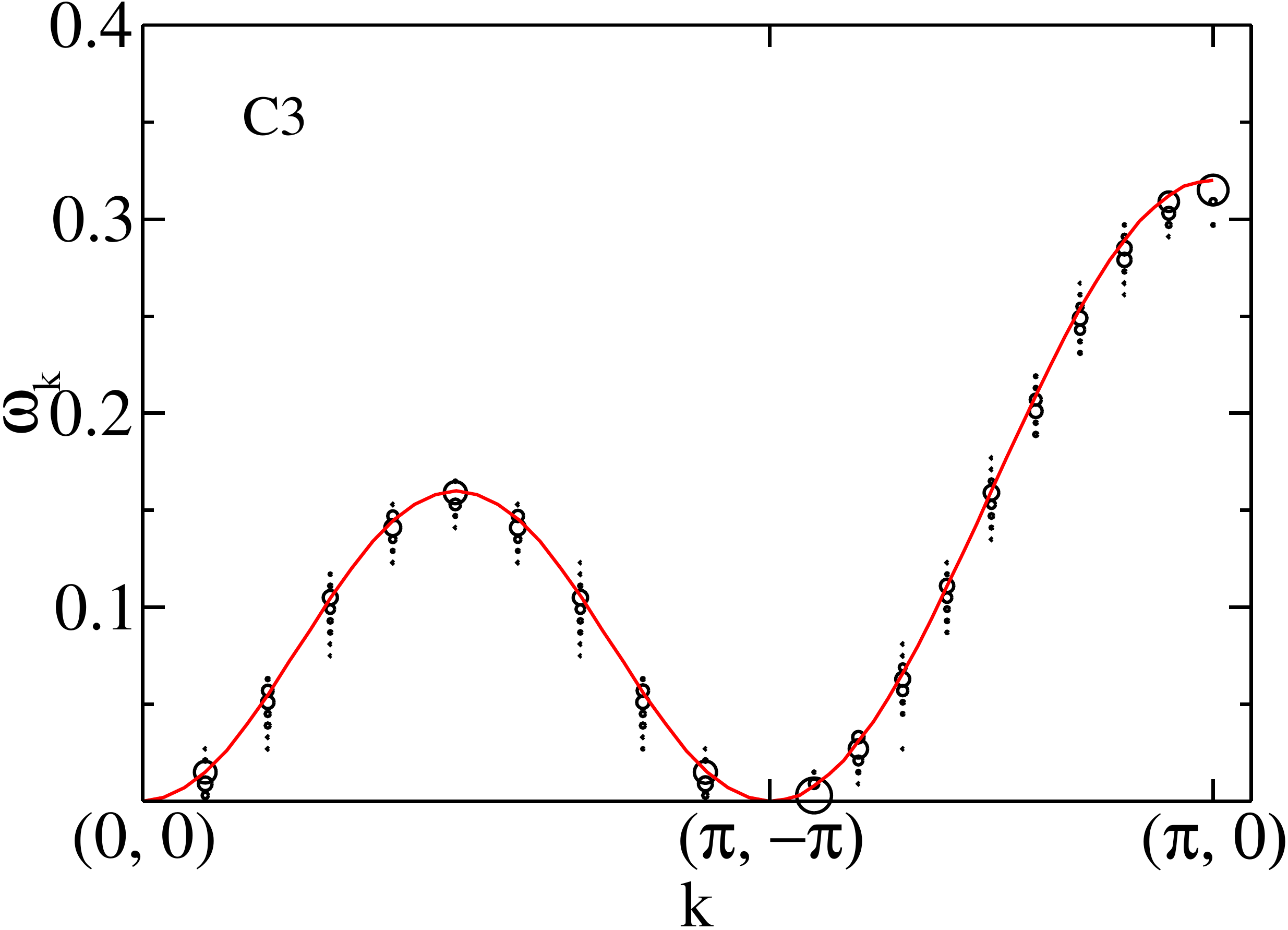}
\includegraphics[width=4.0cm,height=4cm,angle=0,clip=true]{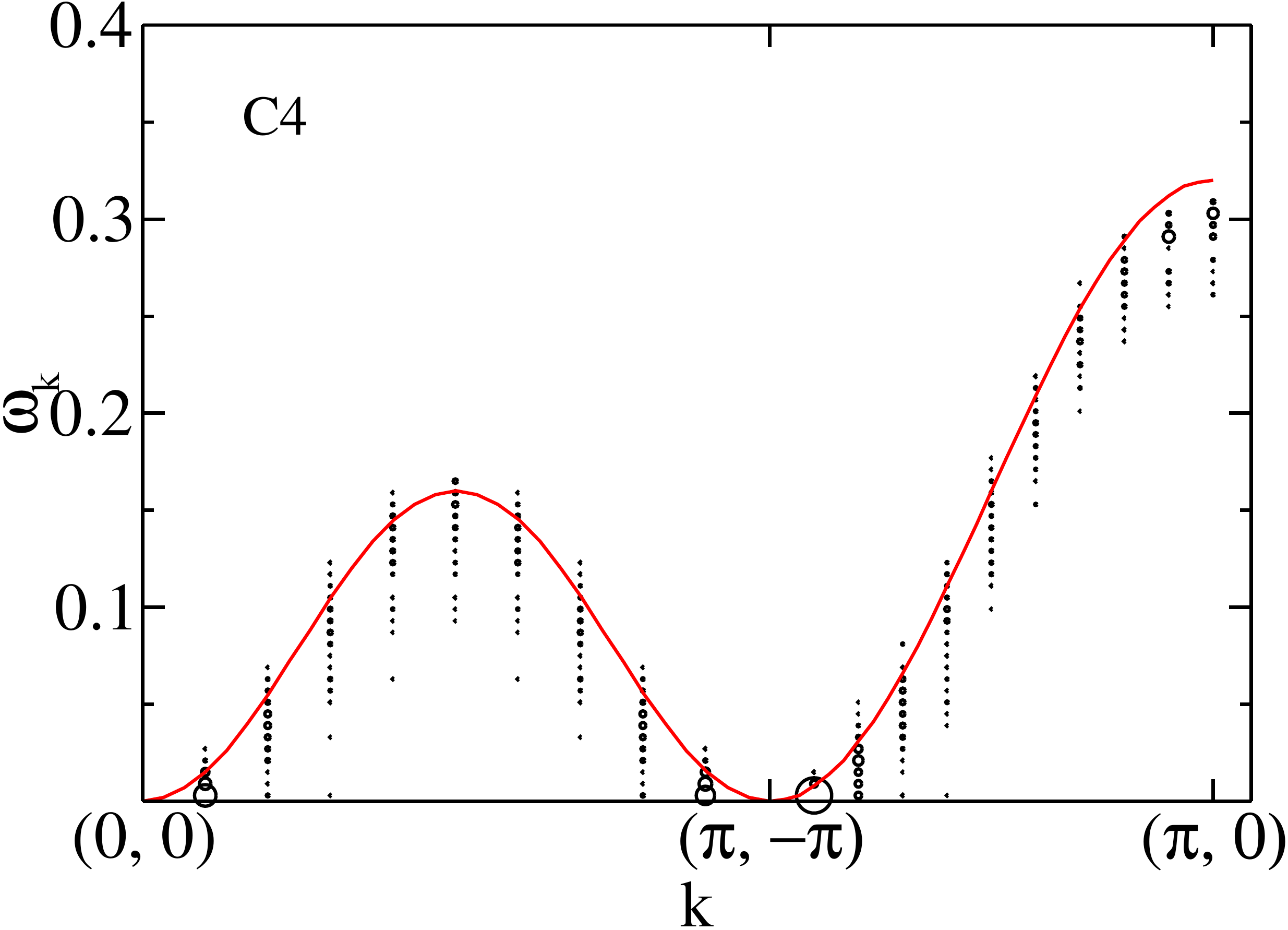}
}
\caption{ (Colour online)
Spin-wave spectra along main symmetry directions of the Brillouin zone
for spin configurations C1, C2, C3 and C4 shown in Fig.1 
($x$ = 0.01, 0.11, 0.21 and 0.41 respectively).  
Increasing fractional weakly coupled
domain boundary spins from C1 to C4
enhances the SW softening near the ZB along $[\pi,0]$ and the spectrum also
becomes broader  for a given ${\bf k}$.
Here $J_{F} =-0.04$, $J_{AF} = 0$, and lattice size is $40 \times 40$.}
\label{fig:swe1}
\end{figure}
\begin{figure}[t]
\centerline{
\includegraphics[width=5.5cm,height=4.5cm,angle=0,clip=true]{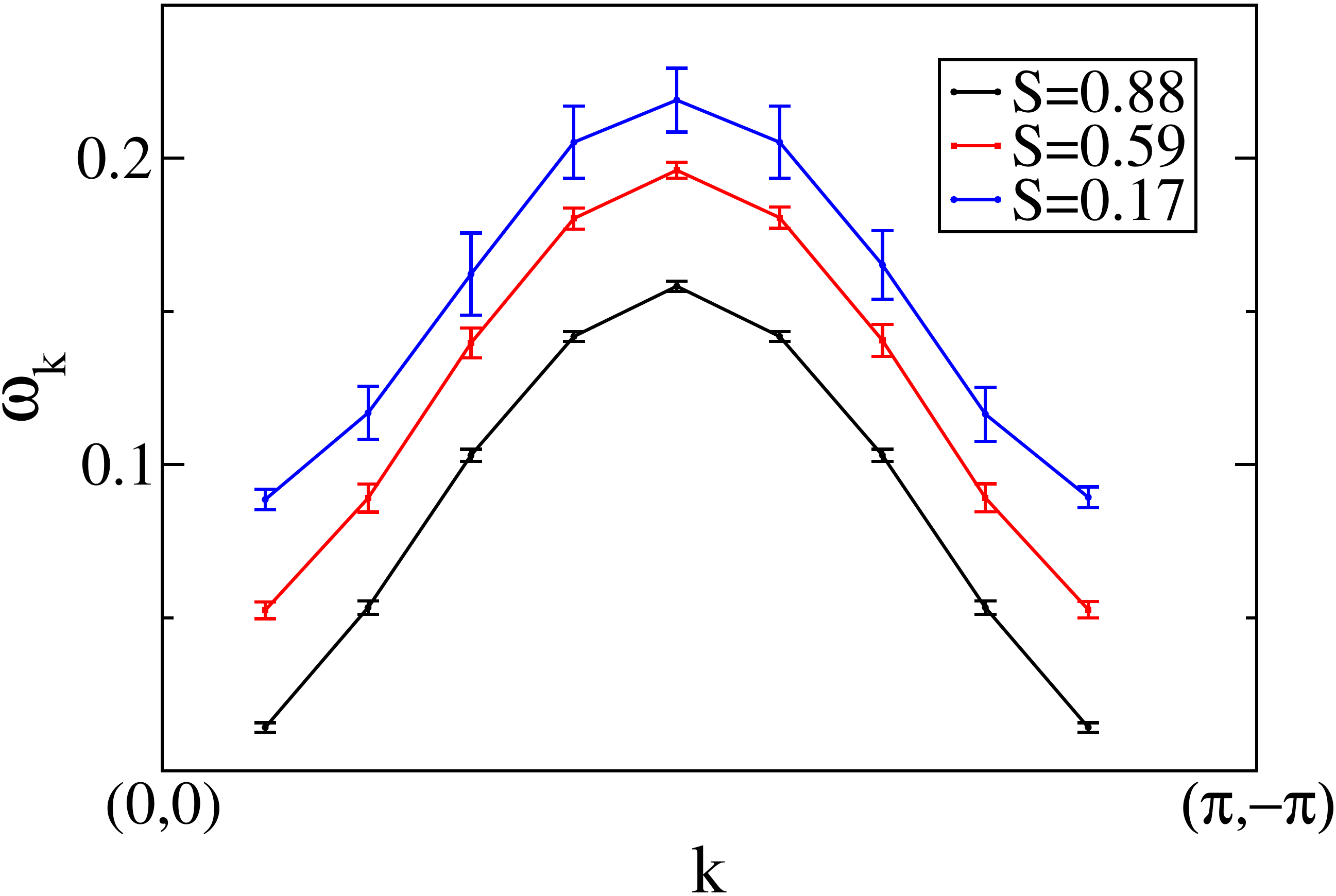}
}
\caption{ (Colour online)  
Mean spin wave energy ${\bar \omega}_{\bf k}$ (dots) and 
spin-wave width $\Delta \omega_{\bf k}$ (bars)
for C2-C4 now with $J_{AF}=0$, {\it i.e}, decoupled domains. 
The curves are vertically shifted for clarity.
}
\end{figure}

\subsection{Broadening: impact of domain size}

There are two ingredients responsible for the spectrum that
one observes in Fig.3, (i)~the domain structure, and 
(ii)~the AF coupling across the domains. To deconvolve these effects
and have a strategy for inferring domain size from neutron data,
we studied a situation where  we set $J_{AF}=0$ in the
Heisenberg model defined on the structures C1-C4.
In that case we will have {\it decoupled FM domains} without
any antiparallel spin orientation between them. We think this
is a interesting scheme to explore since the AF bonds are
limited to the domain boundaries and {\it is not} equal to the
number of mislocated sites.

Fig.5 shows the overall magnon spectrum for this case, using
the same convention as in Fig.3, while Fig.6 quantifies the
mean energy and broadening in this `decoupled domain' case.
The absence of $J_{AF}$ does not seem to make a significant
difference to the spectrum as a comparison of Fig.4 and Fig.6
reveal. 
\begin{figure}[b]
\centerline{
\includegraphics[width=3.3cm,height=3.3cm,angle=0,clip=true]{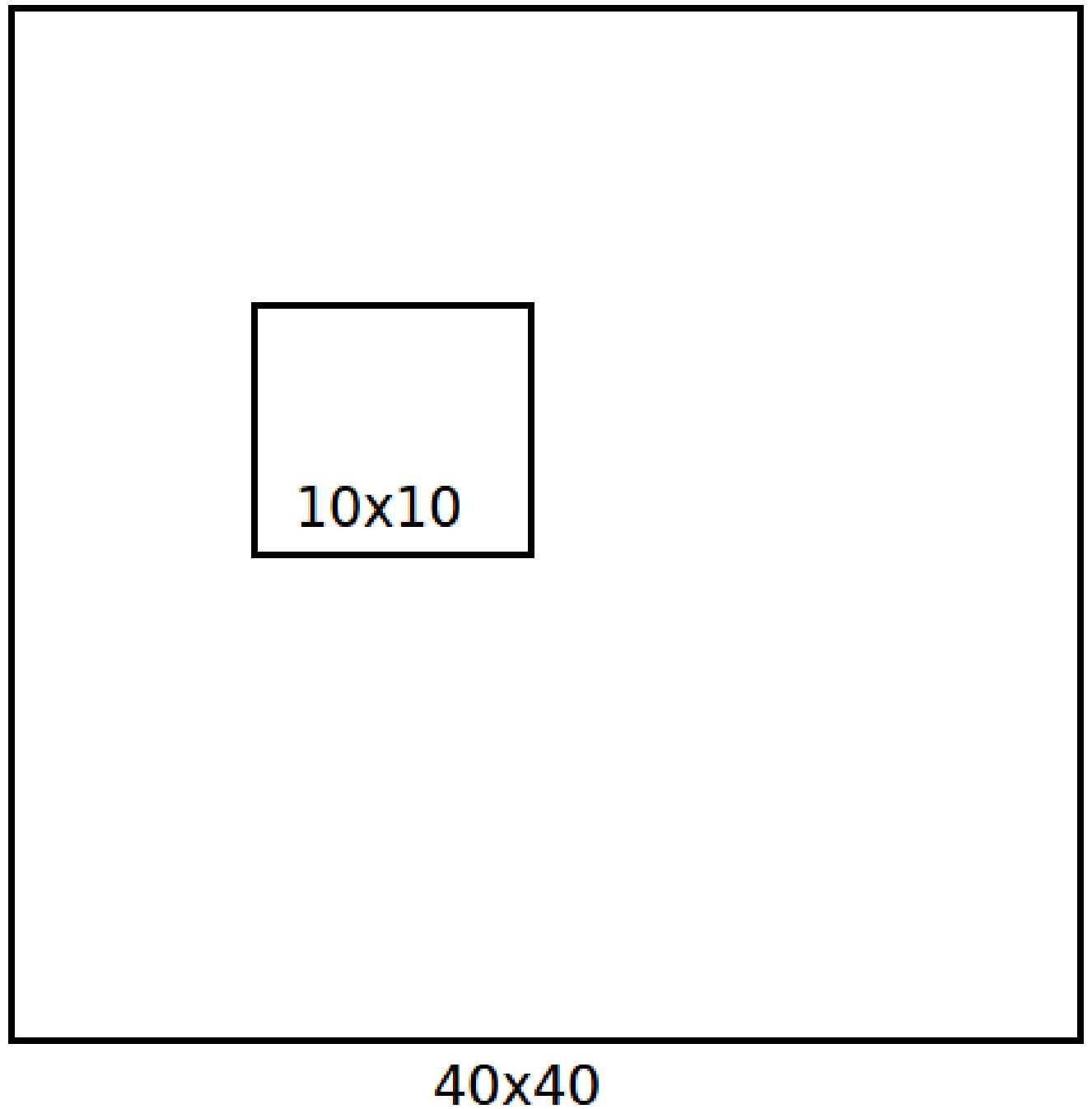}\hspace{0.5cm}
\includegraphics[width=3.3cm,height=3.3cm,angle=0,clip=true]{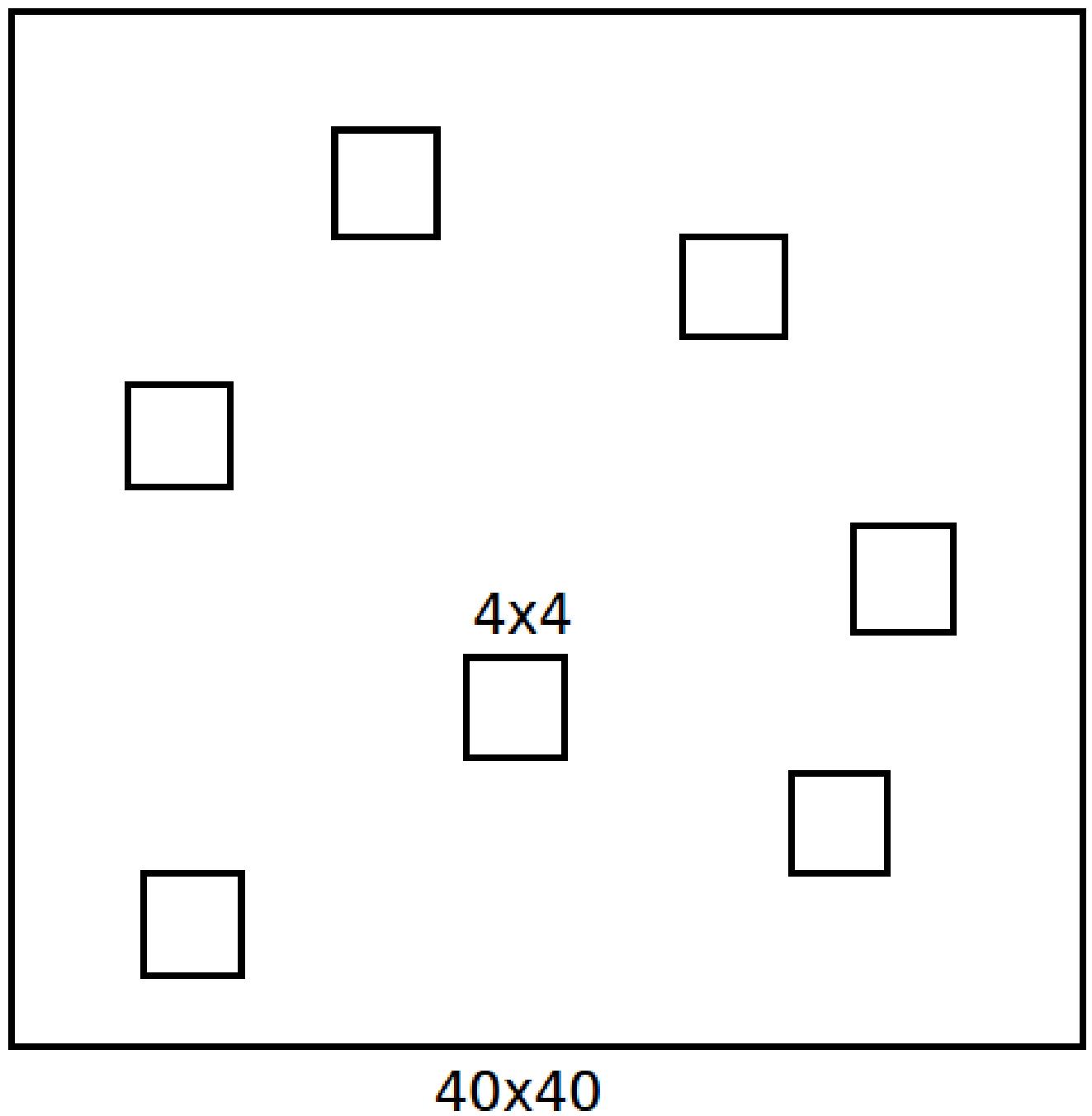}
}
\vspace{.2cm}
\centerline{
\includegraphics[width=3.8cm,height=3.6cm,angle=0,clip=true]{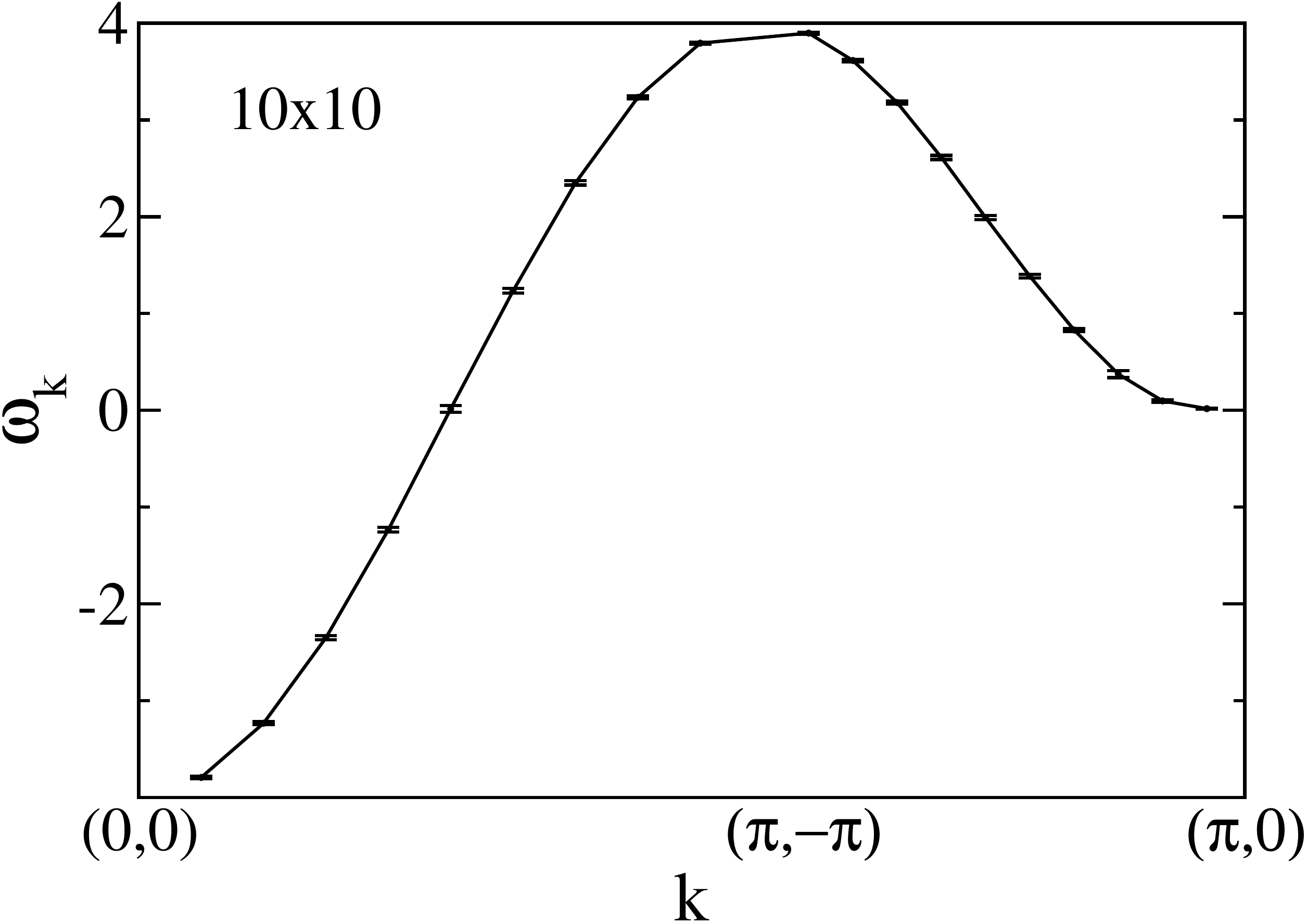}
\includegraphics[width=3.8cm,height=3.6cm,angle=0,clip=true]{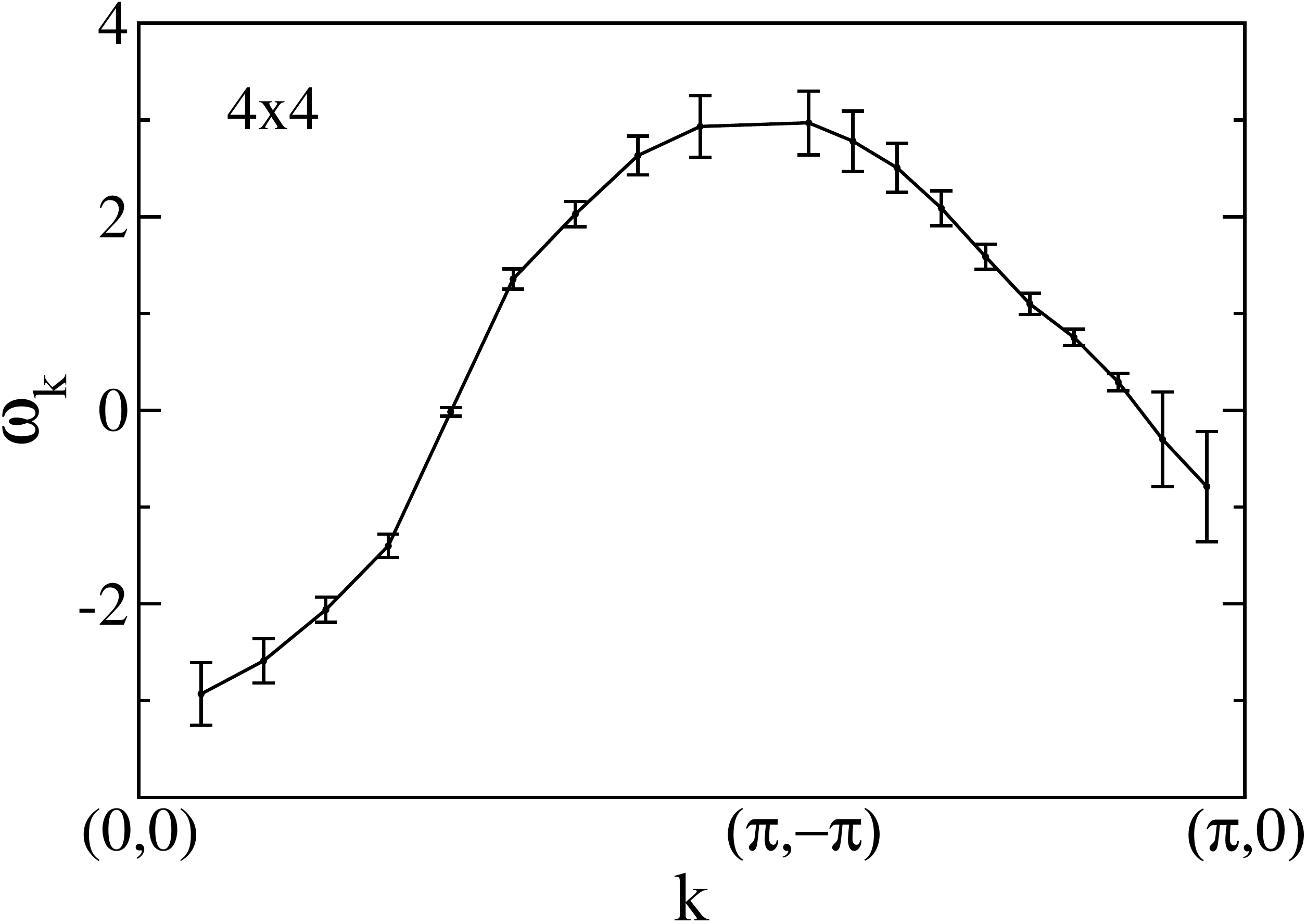}
}
\caption{ (Colour online)  
Modelling C2 in terms of a  domain of size $10 \times 10$ (left)
and of seven domains of size $4 \times 4$ (right). The corresponding
mean energy and broadening are shown below.
}
\end{figure}
This correspondence, valid even in C4, suggests the
following: (i)~most of the spectral features arise from the
domain structure,  and the associated confinement of spin
waves, rather than the AF coupling, and (ii)~we can proceed
with a much simpler modelling of the spectrum and estimation
of domain size without invoking the complicated BdG formulation
that AFM coupling requires.

Essentially, much can be learnt from `tight binding' models
defined on appropriate stuctures, as happens for FM states, 
without having to invoke the `pairing'
terms that arise for AF coupling. A modelling of the 
full dispersion will require the AF terms as well, but
the inference about presence of domains, and an estimate of
their size, need not. 
We proceed with this next.

To make some progress in estimating the typical domain size
we need a few assumptions; (i)~the total degree of mislocation,
$x$, 
should be known, based on the bulk magnetisation measurement.
(ii)~If the overall system size in $L \times L$ (or equivalent
in a 3D model), the number of mislocated sites would be
$x L^2$. (iii)~If the domain size is $L_d$ then the number 
of domains within the $L \times L$ area is
$N_d \sim xL^2/L_d^2$. In reality domains need not
have one single size, as C2-C4 indicate, but we need the
assumption to make some headway.
(iv)~We need to locate these $N_d$ domains randomly, in
a non overlapping manner, within the $L \times L$ system,
and average the spectrum obtained over different realisations
of domain location.

This scheme, carried out for various $L_d$, can be compared
to the full $S({\bf k}, \omega)$ data to get a feel for 
the appropriate $L_d$.
We show the result below for such a tight binding exploration
for the C2 configuration, modelled in terms of different domain
distributions that respect the same overall mislocation.
 
When we compare the ratio of mean broadening to bandwidth 
obtained at different values of $L_d$ (and so $N_d$) with
that for the real data, Fig.4, it turns out that $L_d
=10$ provides a best estimate.
It also reasonably describes the broadening at stronger
disorder, C3 and C4, where of course $N_d$ is larger.
An analytic feel for these results can be obtained by
considering the modes of a square size $L_d \times L_d$
under open boundary conditions.

\subsection{Contrast with uncorrelated antisites}

In modelling the antisite disorder much of the earlier
work in the field assume the 
defect locations to be random. 
We have followed the experimentally motivated path which
suggests that the mislocated sites themselves form an
ordered structure separated from the parent (or majority)
by an antiphase boundary. The sources of scattering are
the boundary between these domains rather than 
random point defects.
Since much of double perovskite modelling has assumed
the random antisite situation it is worth exploring the
differences in the magnon spectrum between correlated
and uncorrelated antisites.

We have already seen the results for correlated disorder
for different degrees of mislocation, $x$. We generated 
{\it uncorrelated} antisite configurations with the same
$x$ by starting with ordered configurations and randomly
exchanging B and B' till the desired degree of disorder
is reached. These configurations naturally do not have any
structural domains. Annealing the full electronic 
Hamiltonian on these configurations, call them $C'_1,~C'_2,..$,
{\it etc}, down to low $T$, leads to the magnetic ground 
states. The ground states are disordered ferromagnets
but without any domain pattern. We computed the magnon
lineshape in these configurations, and, for illustration,
show the results for $C'_2$ and $C'_3$ side by side with
their correlated counterparts $C_2$ and $C_3$.

There is a {\it striking increase} in the magnon line
width (or $\Delta {\omega}_{\bf k}$) in the uncorrelated
case. There is almost nine fold increase in the magnon line
width in $C_2$ and six fold in $C_3$ of the uncorrelated 
disorder with respect to the correlated disorder case. 

\begin{figure*}[t]
\centerline{
\includegraphics[width=2.7cm,height=2.7cm,angle=0,clip=true]{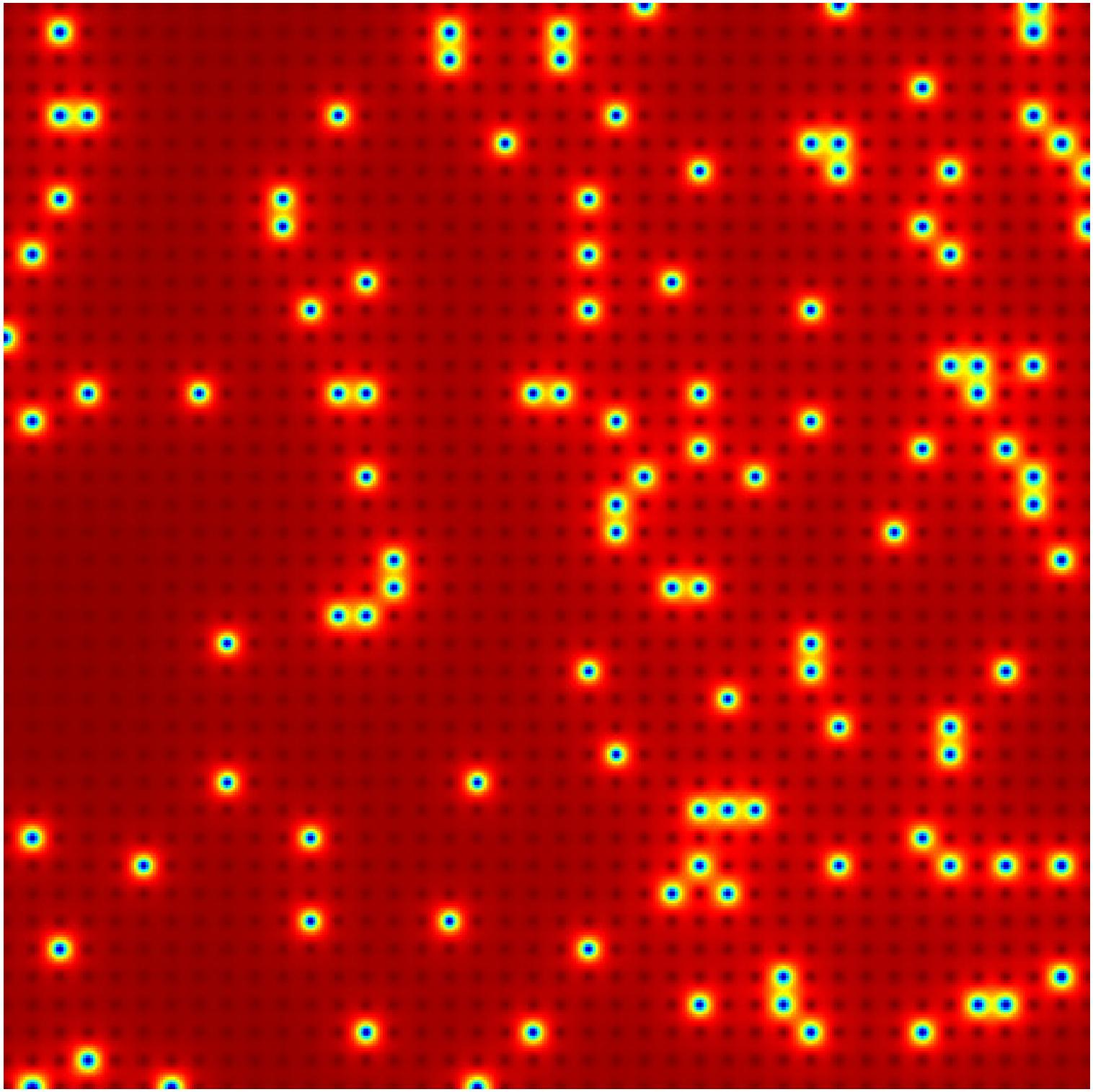}
\includegraphics[width=2.7cm,height=2.7cm,angle=0,clip=true]{C2_str_corr.jpeg}
\hspace{.6cm}
\includegraphics[width=2.7cm,height=2.7cm,angle=0,clip=true]{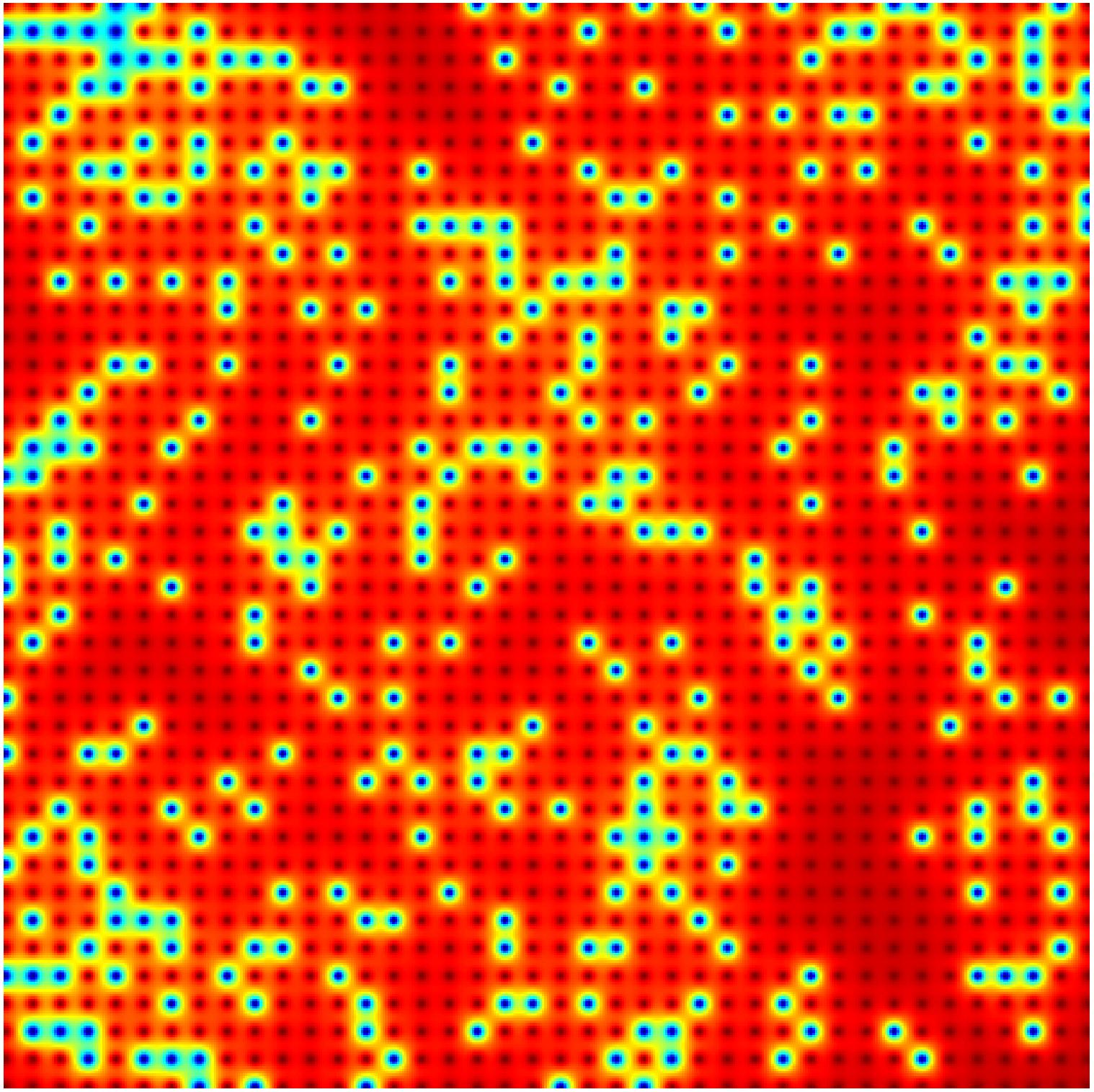}
\includegraphics[width=2.7cm,height=2.7cm,angle=0,clip=true]{C3_str_corr.jpeg}
}
\vspace{.2cm}
\centerline{
\includegraphics[width=2.7cm,height=2.7cm,angle=0,clip=true]{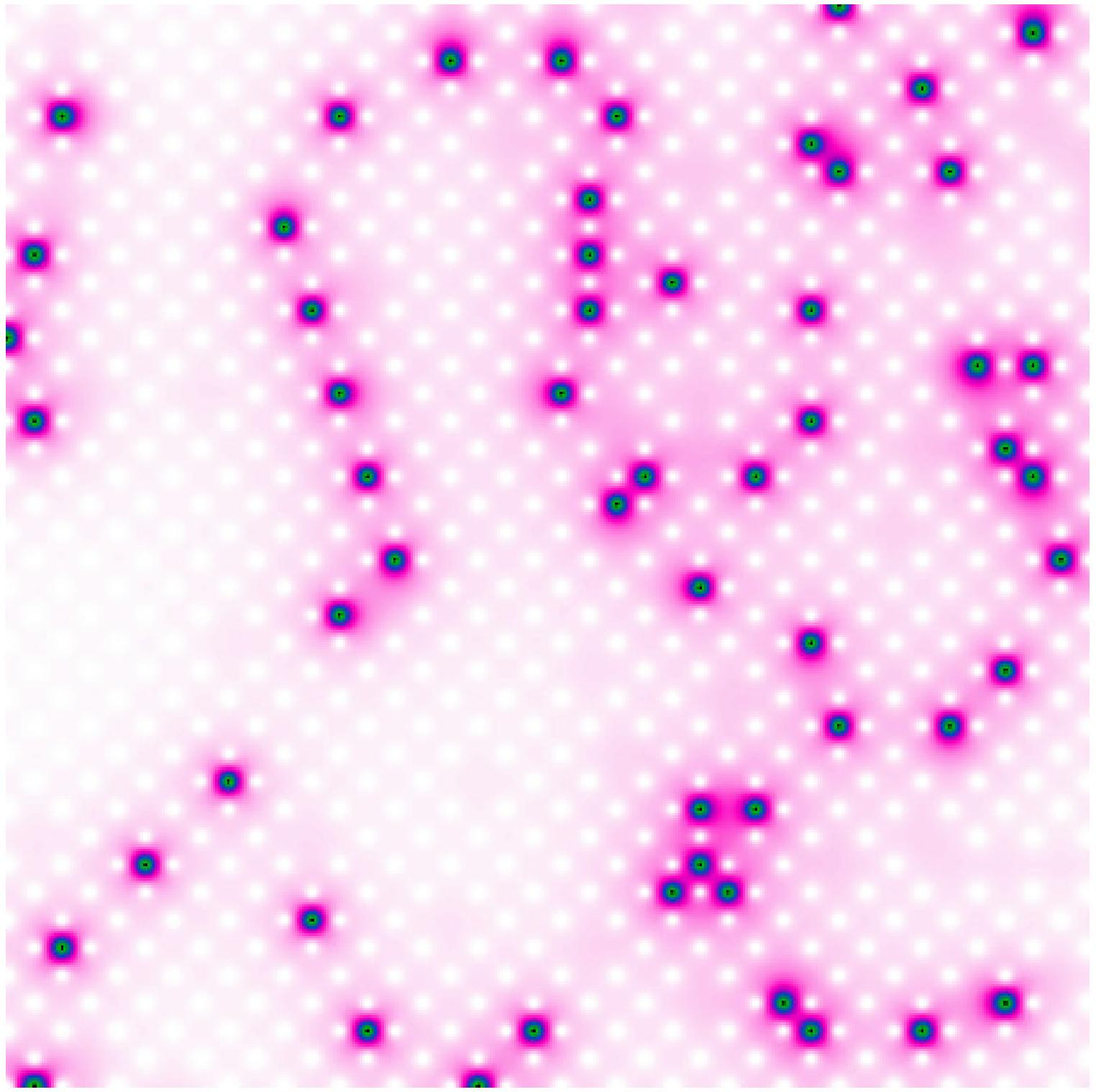}
\includegraphics[width=2.7cm,height=2.7cm,angle=0,clip=true]{C2_spin_corr.jpeg}
\hspace{.6cm}
\includegraphics[width=2.7cm,height=2.7cm,angle=0,clip=true]{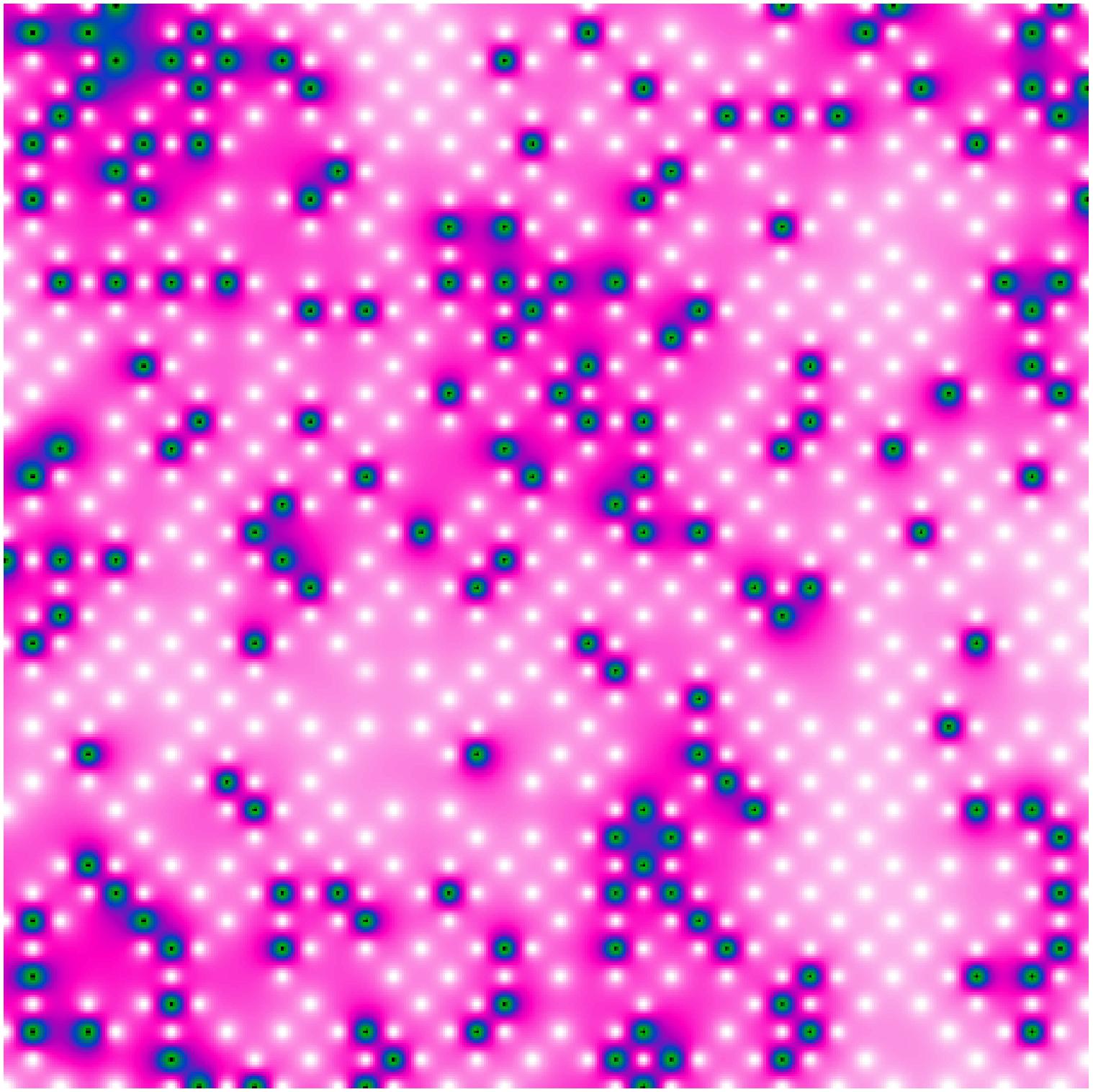}
\includegraphics[width=2.7cm,height=2.7cm,angle=0,clip=true]{C3_spin_corr.jpeg}
}
\vspace{.2cm}
\centerline{
\includegraphics[width=2.9cm,height=2.9cm,angle=0,clip=true]{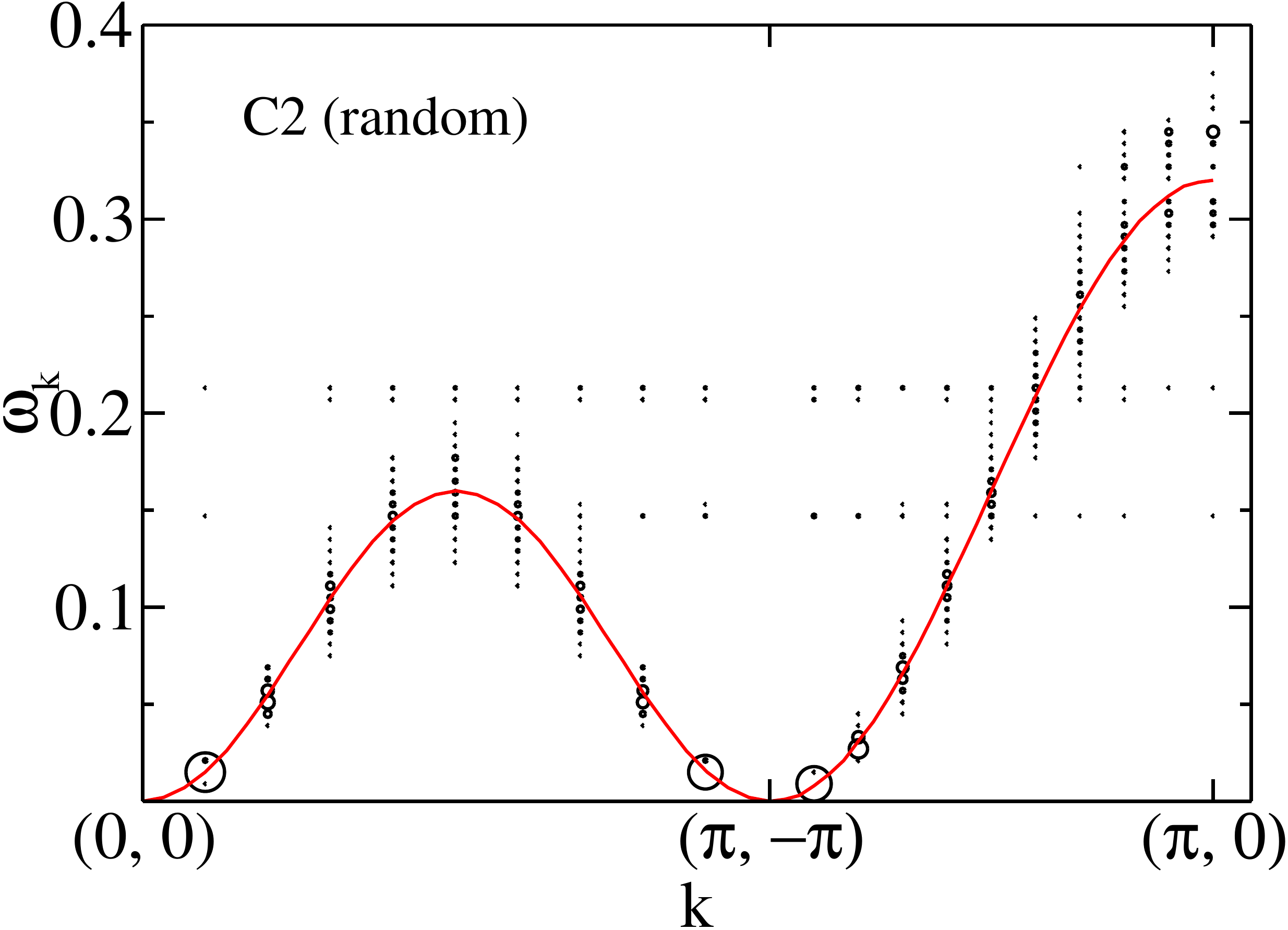}
\includegraphics[width=2.9cm,height=2.9cm,angle=0,clip=true]{C2_SW_corr.pdf}
\hspace{.2cm}
\includegraphics[width=2.9cm,height=2.9cm,angle=0,clip=true]{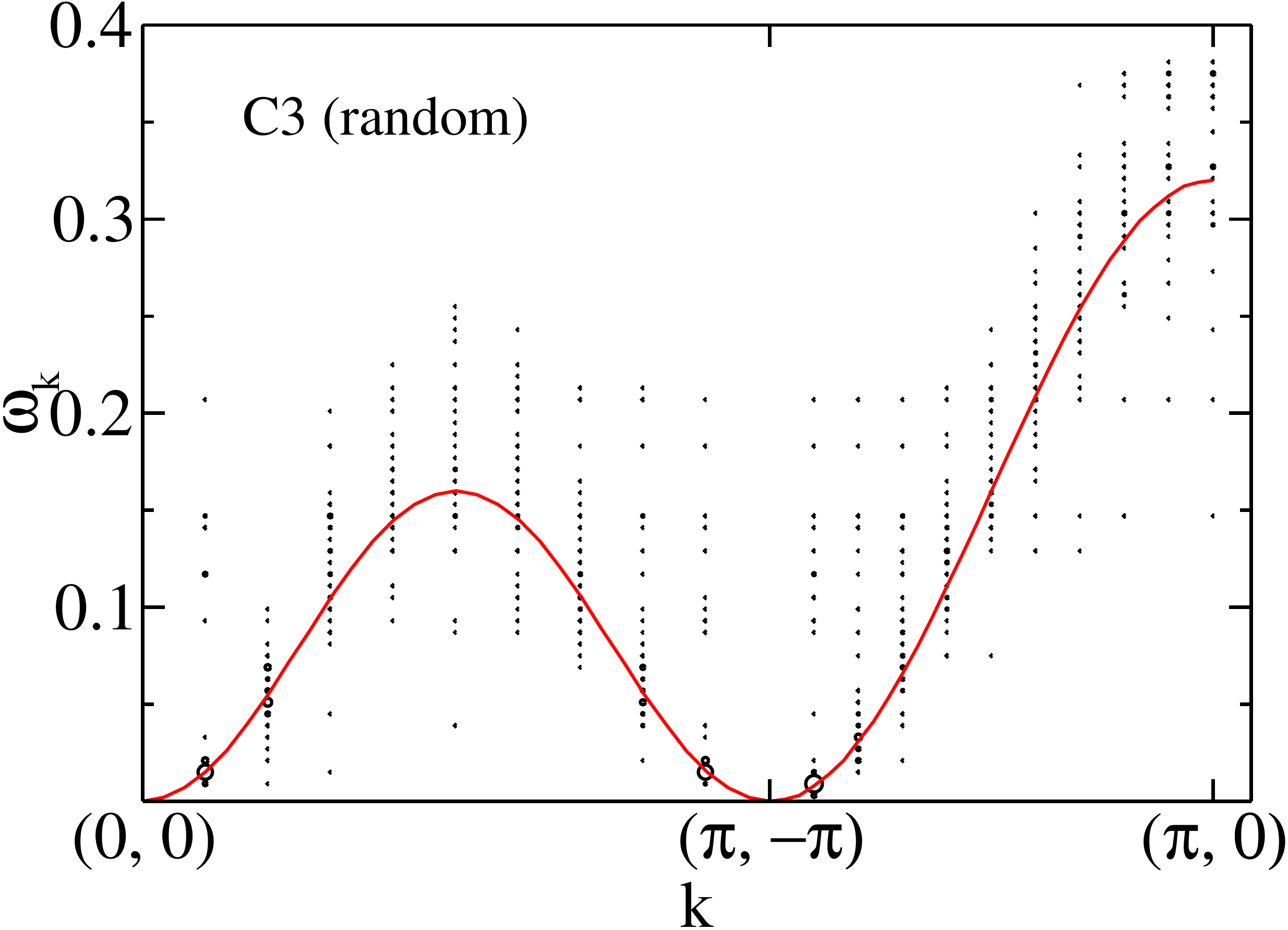}
\includegraphics[width=2.9cm,height=2.9cm,angle=0,clip=true]{C3_SW_corr.pdf}
}
\caption{The left set of panels correspond to mislocation $x=0.11$
where we compare the magnon spectrum for uncorrelated disorder (left)
with correlated disorder (C2) right. the top panels refer to the
structural pattern, the middle to the magnetic ground state, and
the bottom to the magnon response. The right set of panels refer
to $x= 0.21$, and the same indicators as for the left panels.
Notice the remarkably broader lineshape for the uncorrelated
disorder case where it is difficult to make much of a 
correspondence with the clean dispersion.
}
\end{figure*}

\section{Conclusion}
We have studied the dynamical magnetic structure factor of a 
double perovskite system taking into account the basic 
ferromagnetic ordering tendency and the defect induced
local antiferromagnetic correlations. We used structural 
motifs that correspond to correlated disorder, obtained from
an annealing process. The results on magnon energy and broadening
reveal that even at very large disorder, the existence of domain
like structure ensures that the response has a strong similarity
to the clean case. We tried out a scheme for inferring the
domain size from the spin wave damping, so that 
experimenters can make an estimate of domains without having
spatial data, and find it to be reasonably successful. We
also highlight how the common assumption about random antisites,
that is widely used in modelling these materials, would lead
to a gross overestimate of magnon damping. In summary, dynamical
neutron scattering can be a direct probe of the unusual 
ferromagnetic state in these materials and confirm the presence
of correlated antisites.

\section{Acknowledgement}
We acknowledge use of the High Performance Computing facility at HRI.
PM thanks the DAE-SRC Outstanding Research Investigator grant, and the
DST India (Athena) for support.

\section{appendix}
The rotation coefficients are
\be
\begin{array}{c}
p_i^{\pm} = \sqrt{\frac{S}{2}}{(U_i^{xx} \pm U_i^{yy}) - i(U_i^{yx} \mp U_i^{xy})} \\ 
q_i^{\pm} = \sqrt{\frac{S}{2}}{(U_i^{xx} \mp U_i^{yy}) + i(U_i^{yx} \pm U_i^{xy})} \\ 
\hspace{-93pt} r_i^{\pm} = U_i^{zx} \pm iU_i^{zy} \\ 
\hspace{-93pt} p_i^z = U_i^{xz} - iU_i^{yz} \\ 
\hspace{-93pt} q_i^z = U_i^{xz} + iU_i^{yz} \\ 
\hspace{-124pt} r_i^z = U_i^{zz}. 
\end{array}
\ee
And the structure factor coefficients are 
\be
\begin{array}{c}
A_{{}^{mn}_{ij}}^{\alpha \beta} = q_i^{\alpha}p_j^{\beta} u_i^{m^*}u_j^n + 
p_i^{\alpha}q_j^{\beta} v_i^{m^*}v_j^n + 
p_i^{\alpha}p_j^{\beta} v_i^{m^*}u_j^n   \\
+ q_i^{\alpha}q_j^{\beta} u_i^{m^*}v_j^n - S\times r_i^{\alpha}r_j^{\beta} 
(u_i^{m^*}u_i^n + u_j^{m^*}u_j^n) \\ 

B_{{}^{mn}_{ij}}^{\alpha \beta} = q_i^{\alpha}p_j^{\beta} v_i^mv_j^{n^*} +
 p_i^{\alpha}q_j^{\beta} u_i^mu_j^{n^*} + 
p_i^{\alpha}p_j^{\beta} u_i^mv_j^{n^*}  \\ 
+ q_i^{\alpha}q_j^{\beta} v_i^mu_j^{n^*} - S\times r_i^{\alpha}r_j^{\beta}
 (v_i^mv_i^{n^*} + v_j^mv_j^{n^*}). 
\end{array}
\label{GammaDelta}
\ee




\begin{thebibliography}{99}

\bibitem{dp-rev} For reviews, see
D. D. Sarma, Current Op. Solid St. Mat. Sci.,{\bf 5}, 261 (2001),
D. Serrate, J. M. de Teresa and M. R. Ibarra, J. Phys. Cond. Matt. {\bf 19}, 023201 (2007).

\bibitem{nat-kob} K.-I. Kobayashi, T. Kimura, H. Sawada, K. Terakura and Y. Tokura, 
Nature {\bf 395}, 677 (1998).

\bibitem{tom-cryst} 
Y. Tomioka, T. Okuda, Y. Okimoto, R. Kumai, K.-I. Kobayashi, and Y. Tokura, 
Phys. Rev. B {\bf 61}, 422 (2000).

\bibitem{asaka-asd-dom} T. Asaka, X. Z. Yu, Y. Tomioka, Y. Kaneko, T. Nagai,
K. Kimoto, K. Ishizuka, Y. Tokura, and Y. Matsui,
Phys Rev B {\bf 75}, 184440 (2007).

\bibitem{dd-asd-dom} C. Meneghini, Sugata Ray, F. Liscio, F. Bardelli, S. Mobilio, 
and D. D. Sarma,
Phys. Rev. Lett. {\bf 103}, 046403 (2009).

\bibitem{hennion} M. Hennion, et al, Phys. Rev. Lett.  {\bf 94}, 057006 (2005). 

\bibitem{petit} S. Petit, et al, Phys. Rev. Lett.  {\bf 102}, 207201 (2009).

\bibitem{pm-fm-as} V. N. Singh and P. Majumdar, 
Europhys. Lett.  {\bf 94},  47004 (2011).

\bibitem{ps-pm-asd} P. Sanyal, S. Tarat, and P. Majumdar, 
Eur. Phys. J. {\bf B  65}, 39 (2008).

\bibitem{tca} S. Kumar and P.Majumdar, 
Eur. Phys. J. {\bf B 50}, 571-579 (2006).

\bibitem{fishmanJPCM21} R. S. Fishman, 
J. Phys. Cond. Mat. {\bf 21}, 216001 (2009).

\bibitem{hendriksen} P. V. Hendriksen, S. Linderoth, and  P.-A. Lindg\aa{}rd,
J. Phys. Cond. Mat. {\bf 5}, 5675 (1993), Phys. Rev. B {\bf 48}, 7259 (1993).

\end{thebibliography}
\end{document}